\documentclass[aps,twocolumn,preprintnumbers,prd,showpacs,amsmath,amssymb,10pt,nofootinbib,superscriptaddress]{revtex4-2}
\usepackage{graphicx}
\usepackage{scrextend}
\usepackage{manfnt,pifont}
\usepackage{amsmath}
\usepackage{url}
\usepackage{lipsum}
\usepackage[font=small]{caption}
\usepackage[T1]{fontenc}
\usepackage[utf8]{inputenc}
\usepackage{hyperref}
\hypersetup{
  linktocpage,
  colorlinks  = true, 
  urlcolor    = violet, 
  linkcolor   = blue, 
  citecolor   = purple 
}

\usepackage{tikz}
\usepackage{subfigure}
\usepackage{enumitem}
\setlist[itemize]{leftmargin=*}

\newcommand{\bea}{\begin{eqnarray}}
\newcommand{\eea}{\end{eqnarray}}
\newcommand{\be}{\begin{eqnarray}}
\newcommand{\ee}{\end{eqnarray}}

\newcommand{\bma}{\begin{matrix}}
\newcommand{\ema}{\end{matrix}}

\def\beq{\begin{equation}}
\def\eeq{\end{equation}}

\def\cC{{\cal C}}

\def\cG{{\cal G}}

\def\cK{{\cal K}}

\def\cO{{\cal O}}

\def\mA{\mathfrak{A}}
\def\mB{\mathfrak{B}}
\def\mC{\mathfrak{C}}

\def\mS{\mathfrak{S}}

\def\mt{\mathfrak{t}}

\def\ZZ{{\mathbb Z}}

\def\vI{{\vec{I}}}

\def\vmu{{\vec{\mu}}}

\def\vh{{\vec{h}}}

\def\Im{{\rm Im \,}}

\def\det{{\rm det \,}}

\def\half{{1\over 2}}

\def\p{\partial}

\def\om{\omega}

\def\xx{{\boldsymbol x}}

\def\ki{\chi}

\def\pa{{p_a}}
\def\pb{{p_b}}
\def\Ber{{\rm Ber}}
\def\no{\nonumber}
\def\sm{\smallskip}
\def\mgh{\mathcal{G}}

\DeclareMathAlphabet\mathbfcal{OMS}{cmsy}{b}{n}

\makeatletter
\protected\def\onecolumngrid{%
  \do@columngrid{one}{\@ne}%
  \let\set@footnotewidth\set@footnotewidth@one
}
\makeatother

\begin{document}

\preprint{UUITP--18/26}

\title{Degenerations of flat connections on Riemann surfaces}
\author{Mattia Biancotto}
\email{mattia.biancotto@aei.mpg.de}
\affiliation{Max-Planck-Institut f\"ur Gravitationsphysik (Albert-Einstein-Institut)
Am M\"uhlenberg 1, 14476 Potsdam, Germany}

\author{Eric D'Hoker}
\email{dhoker@physics.ucla.edu}
\affiliation{Mani L. Bhaumik Institute for Theoretical Physics, Department of Physics and Astronomy, University of California, Los Angeles, CA 90095, USA}

\author{Axel Kleinschmidt}
\email{axel.kleinschmidt@aei.mpg.de}
\affiliation{Max-Planck-Institut f\"ur Gravitationsphysik (Albert-Einstein-Institut)
Am M\"uhlenberg 1, 14476 Potsdam, Germany}

\author{Michele Santagata}
\email{michele.santagata@physics.uu.se}
\affiliation{Department of Physics and Astronomy, Uppsala University, 75120 Uppsala, Sweden}

\author{Oliver Schlotterer}
\email{oliver.schlotterer@physics.uu.se}
\affiliation{Department of Physics and Astronomy, Uppsala University, 75120 Uppsala, Sweden}
\affiliation{Department of Mathematics, Centre for Geometry and Physics, Uppsala University, 75120 Uppsala, Sweden}

\begin{abstract}

\noindent The integration kernels for polylogarithm functions on a compact Riemann surface of arbitrary genus $h$ are shown to close as the surface undergoes a non-separating degeneration to one of genus $h{-}1$. Explicit formulas are obtained for the non-separating degeneration of the multivariable Enriquez connection for genus $h$ with an arbitrary number of variables to the Enriquez connection for genus $h{-}1$ with two additional punctures whose Lie algebra generators are related to the original ones by the characteristic Bernoulli generating functions known from the degeneration at $h=1$. Analogous degeneration formulas are obtained for the single-valued DHS kernels at the leading order in the real degeneration parameter that is adapted to relating modular tensors at genus $h$ and $h{-}1$.
\end{abstract}
\maketitle

\section{Introduction} 

Flat connections valued in infinite-dimensional Lie algebras provide powerful tools for generating spaces of  homotopy-invariant iterated integrals that form an algebra under addition and multiplication and close under taking derivatives and primitives. Flat connections on Riemann surfaces of genus zero produce standard polylogarithms \cite{Lappo:1953, Goncharov:1995}, while their genus one counterparts  lead to elliptic polylogarithms in one variable \cite{Levin:1997, CEE, LR, Enriquez:2023} or  in multiple variables \cite{CEE,Brown:2011wfj}. These spaces are enjoying an increasingly central role in organizing the structure of perturbative amplitudes in quantum field theory \cite{Weinzierl:2022eaz, Bourjaily:2022bwx} and string theory \cite{Berkovits:2022ivl, DHoker:2024xor}. 

On a compact Riemann surface $\Sigma$ of arbitrary genus $h \geq 1$, the Enriquez connection in one or in several variables \cite{Enriquez:2011} is meromorphic in its variables and in the moduli of $\Sigma$,  but it fails to be single-valued or to be modular invariant. The DHS connection introduced  in \cite{DHoker:2023vax} for one variable  and in  \cite{DHoker:2026lgg} for multiple variables is single-valued and modular invariant, but fails to be meromorphic.

The Enriquez and DHS connections  with the same number of variables on the same surface $\Sigma$ are related to one another by the composition of a gauge transformation and an automorphism of the Lie algebra in which they take values \cite{DHoker:2025szl,DHoker:2026lgg}. One or several of the variables in a connection may be traded for one or several  punctures by freezing the position of the corresponding variables and reducing the Lie algebra accordingly \cite{DHoker:2026lgg}. In either case, the connection may be extended to a connection in which the moduli of the Riemann surface $\Sigma$ are allowed to vary while the topology of $\Sigma$ is kept fixed. For genus one, the meromorphic universal connection was developed in \cite{CEE} while for higher genus the global extension of the DHS connection is under investigation  in~\cite{DESZ:2026}.

Each one of the above-mentioned connections is attached to a surface of given topology. In the present paper, we pursue a program whose goal is to relate the connections and eventually the iterated integrals on surfaces of different topologies to one another in an analytically controlled manner. Beyond its intrinsic mathematical interest, this avenue of research is motivated by several applications in theoretical physics. In the context of string perturbation theory, such links between Riemann surfaces of different genera provide valuable tools to check or exploit unitarity and to systematically compute coefficients in the low-energy expansion of scattering amplitudes \cite{DHoker:2017pvk, DHoker:2018mys, Pioline:2018pso, DHoker:2019blr, Edison:2021ebi, Eberhardt:2022zay}. In quantum field theory, the differential-equation approach to Feynman integrals already mixes different geometries within a single integral family which makes the control of degeneration limits of the associated connections even more pressing \cite{Delto:2023kqv, Driesse:2024feo, Duhr:2024bzt, Coro:2025vgn, Becchetti:2025rrz}. 

The topology of a surface  $\Sigma$ of genus $h \geq 1$  can change as a result of the degeneration of a homotopically non-trivial curve $\mC$ of the surface. One distinguishes two cases.  A~\textit{separating degeneration} occurs when $\mC$ belongs to the trivial homology class and the surface separates into two connected surfaces of genera $h_1+h_2=h$. A~\textit{non-separating degeneration} occurs when $\mC$ is homologically non-trivial and $\Sigma$ degenerates into a surface of genus $h-1$ with two punctures, denoted by $p_a$, $p_b$ in the figure below.

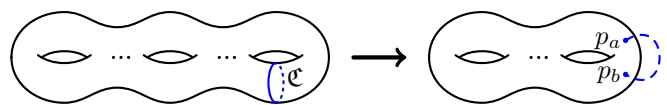
\vspace{-0.1cm}
\begin{center}
\begin{tikzpicture}[xscale=.5, yscale=.5, line cap=round, line join=round]

\begin{scope}[xshift=0.3cm, yshift=0cm]
        \draw[thick] (-4.2,0) to[out=90,in=180] (-2.8,1.2) 
                             to[out=0,in=180] (-1.4,0.8)
                             to[out=0,in=180] (0,1.2)
                             to[out=0,in=180] (1.4,0.8)
                             to[out=0,in=180] (2.8,1.2)
                             to[out=0,in=90] (4.2,0)
                             to[out=-90,in=0] (2.8,-1.2)
                             to[out=180,in=0] (1.4,-0.8)
                             to[out=180,in=0] (0,-1.2)
                             to[out=180,in=0] (-1.4,-0.8)
                             to[out=180,in=0] (-2.8,-1.2)
                             to[out=180,in=-90] (-4.2,0);

        \draw[thick] (-3.5, 0.06) .. controls (-3.1, -0.22) and (-2.5, -0.22) .. (-2.1, 0.06);
        \draw[thick] (-3.4, 0.00) .. controls (-3.1,  0.22) and (-2.5,  0.22) .. (-2.2, 0.00);

        \fill (-1.5,0) circle (1.2pt);
        \fill (-1.3,0) circle (1.2pt);
        \fill (-1.1,0) circle (1.2pt);

        \draw[thick] (-0.7, 0.06) .. controls (-0.3, -0.22) and (0.3, -0.22) .. (0.7, 0.06);
        \draw[thick] (-0.6, 0.00) .. controls (-0.3,  0.22) and (0.3,  0.22) .. (0.6, 0.00);

        \fill (1.3,0) circle (1.2pt);
        \fill (1.5,0) circle (1.2pt);
        \fill (1.7,0) circle (1.2pt);

        \draw[thick] (2.1, 0.06) .. controls (2.5, -0.22) and (3.1, -0.22) .. (3.5, 0.06);
        \draw[thick] (2.2, 0.00) .. controls (2.5,  0.22) and (3.1,  0.22) .. (3.4, 0.00);

        \end{scope}

    \newcommand{\y}{-0.7}
\draw[->, ultra thick] (5.15,0) -- (6.55,0);
\draw[thick, dotted, blue] (3.1, -1.2) arc (-90:90:0.2 and 0.52);
        \draw[thick, blue] (3.1, -0.15) arc (90:270:0.2 and 0.52);

    \node at (3.6,-0.5) {$\mC$};

   \begin{scope}[xshift=11.35cm, yshift=0cm]
        \draw[thick] (-4.2,0) to[out=90,in=180] (-2.8,1.2) 
                             to[out=0,in=180] (-1.4,0.8)
                             to[out=0,in=180] (0,1.2)
                             to[out=0,in=90] (1.4,0)   
                             to[out=-90,in=0] (0,-1.2)
                             to[out=180,in=0] (-1.4,-0.8)
                             to[out=180,in=0] (-2.8,-1.2)
                             to[out=180,in=-90] (-4.2,0);

        \draw[thick] (-3.5, 0.06) .. controls (-3.1, -0.22) and (-2.5, -0.22) .. (-2.1, 0.06);
        \draw[thick] (-3.4, 0.00) .. controls (-3.1,  0.22) and (-2.5,  0.22) .. (-2.2, 0.00);

        \fill (-1.5,0) circle (1.2pt);
        \fill (-1.3,0) circle (1.2pt);
        \fill (-1.1,0) circle (1.2pt);

        \draw[thick] (-0.7, 0.06) .. controls (-0.3, -0.22) and (0.3, -0.22) .. (0.7, 0.06);
        \draw[thick] (-0.6, 0.00) .. controls (-0.3,  0.22) and (0.3,  0.22) .. (0.6, 0.00);


        \coordinate (pa1) at (1, 0.45);
        \coordinate (pb1) at (1, -0.45);
        \fill[blue] (pa1) circle (2pt) node[left, black, xshift=2pt] {$p_{a}$};
        \fill[blue] (pb1) circle (2pt) node[left, black, xshift=2pt] {$p_{b}$};
        \draw[blue, thick, dashed] (pa1) .. controls (2.2, 1.2) and (2.2, -1.2) .. (pb1);
    \end{scope}
\end{tikzpicture}
\end{center}
\sm 

In the remainder of this paper, we shall investigate the non-separating degeneration of the Enriquez and DHS integration kernels and multi-variable connection in the case where a non-trivial homology cycle pinches. 
The main results are \eqref{3.b.1} and \eqref{4.b.1aa} for the degeneration of Enriquez and DHS kernels as a single cycle pinches, \eqref{5.6} for the resulting degeneration of the Enriquez connection and \eqref{h0degen} for its behaviour under sequential degenerations of $\Sigma$ to a multi-nodal sphere after pinching $h$ cycles.
The behaviour under the degeneration of more general cycles follows from modular transformations, which is readily available for our degeneration results on DHS kernels and relegated to future work for the case of Enriquez kernels.


\section{The Enriquez connection} 

\label{sec:2}
Consider a compact Riemann surface $\Sigma$ of genus $h\geq 1$ whose homology group $H_1(\Sigma, \ZZ)$ is equipped with a symplectic intersection pairing. Choose a canonical homology basis of cycles $\mA^I$ and $\mB_I$ for $I=1,\cdots, h$ and dual holomorphic $(1,0)$ forms $\om_I$ whose $\mA$-periods are normalized canonically and whose $\mB$-periods give the period matrix~$\Omega$, 
\bea
\label{2.1}
\oint _{\mA^I } \om_J = \delta^I_J \hskip 1in 
\oint _{\mB_I} \om_J = \Omega_{IJ}
\eea  
The modular group $Sp(2h,\ZZ)$ acts linearly on the cycles $\mA, \mB$ and leaves the intersection pairing invariant. The  Enriquez connection $\cK_\text{E}$ in $n$ variables without punctures  is defined on the configuration space ${\rm Cf}_n(\widetilde \Sigma) = \widetilde \Sigma^n \smallsetminus \{ \hbox{diagonals} \}$ of $n$ distinct points $\xx=(x_1, \cdots, x_n)$ in the universal cover $\widetilde \Sigma$ of $\Sigma$ and takes values in the infinite-dimensional Lie algebra $\mt_{h,n}$
generated by $a_i^I, b_{jJ}, t_{ij}=t_{ji} $ ($i\neq j$) for $I,J=1,\cdots, h$ and $i,j=1,\cdots, n$, subject to the structure relations \cite{Enriquez:2011}, 
\begin{align}
\label{2.2}
{} \big [ a_i^I, a_j^J \big]   = \big [b_{iI}, b_{jJ} \big] & =0 & \big[ b_{iI}, a_j^J \big ] &= \delta ^J_I t_{ij}
\no \\
{} \big [ a_i^I, t_{jk} \big]   = \big [b_{iI}, t_{jk} \big] & =0 & \big[ b_{iI}, a_i^I \big] &= - \sum_{j \not= i} t_{ij} 
\end{align}
where $i, j,k$ are pairwise distinct. The  Enriquez connection $\cK_\text{E}(\xx)$ is a meromorphic one-form whose $\mA$ monodromy in each variable $x_i$ is trivial while its monodromy in $x_i$ along a cycle $\mB_K$ is given by \cite{Enriquez:2011,DHoker:2026lgg}, 
\bea
\label{2.3}
\cK_\text{E}( \mB_{i K} \cdot \xx) = e^{-2 \pi i b_{iK}} \cK_\text{E}(\xx) \, e^{ 2 \pi i b_{iK}}
\eea

There exists a preferred fundamental domain $D_p$ representing $\Sigma$ in $\widetilde \Sigma$ in which the only poles of $\cK_\text{E}(\xx)$ in $x_i$ are simple poles at $x_j$ with residue $-t_{ij}$ for $j \not= i$. Outside of $D_p$, the pole structure is mandated by the monodromy relations. In view of these properties and the flatness conditions, $\cK_\text{E}(\xx)$ admits a Lie series expansion in powers of the generators $b_{iI}$,
\begin{align}
 \cK_\text{E} (\xx)   & =    \sum _\vI \bigg ( \sum _i dx_i  \, \varpi ^\vI{}_J (x_i) \, B_{i \vI} \, a_i ^J \notag \\
&  \qquad  \qquad + \sum_{i\not=j} dx_i \, \chi ^\vI (x_i,x_j) B_{i\vI} \, t_{ij} \bigg ) 
\label{2.4}
\end{align}
where the sum is over all words $\vI = I_1 \cdots I_r$ in the alphabet $I_1, \cdots, I_r \in \{ 1, \cdots, h\}$ for arbitrary length $r\geq 0$, and we have defined the shorthand $B_{i \vI} = B_{i I_1} \cdots B_{i I_r}$ where $B_{iI} = {\rm ad}_{b_{iI}}$. The $(1,0)$ forms $\varpi$ and $\chi$ are the traceless and trace parts of the Enriquez kernels, 
\bea
\label{2.5}
g^{I_1 \cdots I_r}{}_J (x,y) = \varpi ^{I_1 \cdots I_r} {}_J(x) - \chi^{I_1 \cdots I_{r-1}} (x,y) \, \delta ^{I_r}_J
\eea
for arbitrary points $x,y \in  \widetilde \Sigma$, and we set $\varpi ^\emptyset {}_J(x) = \om_J(x)$ in (\ref{2.4}). The properties of $\varpi$, $\chi$ and $g$ are induced by those of $\cK_\text{E}$. In particular, they all have vanishing $\mA$ monodromies; their $\mB$ monodromies may be deduced from (\ref{2.3}) but will not be needed here; $\varpi^{\vI}{}_J(x)$ is holomorphic in $x \in \widetilde \Sigma$;  for $\vI \not= \emptyset$ the one-forms  $\chi^\vI(x,y)$ and $g^{\vI K}{}_J(x,y)$ are holomorphic in the  preferred fundamental domain $D_p$; and $\chi(x,y)$ has a simple pole in $x$ at $y$ with  residue $-1$. The following period integral will come in handy,
\bea
\label{2.5a}
\oint _{\mA^K} dt \, g^{I_1 \cdots I_r}{}_J(t,y) = (-2 \pi i)^r { \Ber_r \over r!} \delta ^{I_1 \cdots I_r K}_J
\eea
where $\Ber_r$ are the Bernoulli numbers and the generalized Kronecker delta is given by $\delta ^{I_1 \cdots I_s}_J = \delta^{I_1}_J \cdots \delta^{I_s}_J$.

\sm

Explicit expressions for the Enriquez kernels $g^\vI{}_J(x,y)$ in terms of functions that are standard in the theory of higher genus Riemann surfaces were advanced only recently in \cite{Enriquez:2021} and \cite{DHoker:2025dhv} (also see \cite{Baune:2024biq} for Poincar\'e-series representations that converge in restricted subsets of the moduli space of compact Riemann surfaces). Theorem 1 of \cite{DHoker:2025dhv} gives the lowest rank case as~follows,
\begin{equation}
\label{2.6}
    g^I {}_J (x,y) = \oint _{\mA^I} dt \, \om_J(t) \, \p_x \ln {E(x,y) \over E(x,t)} - \pi i \, \delta ^I_J \, \om_J(x)
\end{equation}
where $E(x,y)$ is the prime form \cite{Fay}. 
One has the following relation among prime forms and the traceless part of the Enriquez kernel,
\begin{equation}
    \label{2.6a}
\p_x \ln {E(x,z) \over E(x,y)}= \chi(x,y) - \chi(x,z) =  \ki_{yz}(x)
\end{equation}
i.e.\ $\ki_{yz}(x)$ is a meromorphic one-form in $x$, with simple poles at $y$ and $z$ with residues $-1$ and $1$, respectively, and zero $\mA$-periods. 
Theorem 2 of \cite{DHoker:2025dhv} provides a recursion relation by convolution integrals over $\mA$-cycles  for the case of $\vI = I_1 \cdots I_r$ of arbitrary rank,
\bea
\label{2.7}
g^{ K \vI }{}_J(x,y) & = & - \oint _{\mA^K}  dt \, g^L{}_J (x,t) \, g^{\vI}{}_L(t,y)
\no \\ &&
- \sum _{k=1}^{r-1} (- 2 \pi i )^k { \Ber_k \over k!} \delta ^{I_1 \cdots I_k}_K \, g^{K I_{k+1} \cdots I_r}{}_J (x,y)
\no \\ &&
- \om_J(x) (- 2 \pi i)^{r+1} {\Ber_{r+1} \over r!} \delta ^{K \vI }_J 
\eea
The difference of the trace part at points $y,z$ 
\beq
\chi^{\vI}_{yz}(x)  =
\chi^{\vI}(x,y) - \chi^{\vI}(x,z)
\label{chifun}
\eeq
has vanishing $\mA$-periods and obeys the simplified~formula,
\begin{align}
  \chi^{K \vI}_{yz}(x)   &=  \oint _{\mA^K}   dt \, \chi(x,t)  \chi^{\vI}_{yz}(t) \label{2.7a}\\
 &\quad - \sum_{k=1}^r (-2 \pi i)^k { \Ber_k \over k!} \, \delta ^{I_1 \cdots I_k}_K   \chi_{yz}^{K I_{k+1} \cdots I_r}(x)  
\notag
\end{align}
Below, we shall use these formulas as the starting point for the evaluation of the degeneration of the Enriquez connection and kernels. 

\section{Non-separating degenerations}
\label{sec:3}

A non-separating degeneration occurs when a cycle $\mC$ in a nontrivial homology class of the Riemann surface degenerates. To evaluate the non-separating degeneration of a modular invariant function, we may use its $Sp(2h,\ZZ)$-invariance to carry out the degeneration on an arbitrary cycle. The customary choice of the cycle $\mB_h$ for a surface $\widehat \Sigma$ of genus $h$ results in a genus $h{-}1$ surface $\Sigma$ with two additional punctures and reduces the modular group from $Sp(2h,\ZZ)$ to $Sp(2h{-}2,\ZZ)$. Here and in the remainder of this work, we shall employ the notation that a hat indicates quantities associated with the genus $h$ surface while we omit the hat for the genus $h{-}1$ counterparts. Examples are $\widehat \Sigma$ or $\widehat g^{\vI}{}_J(x,y)$ and only when there is a natural embedding such as for the homology cycles we will drop this notational distinction.

For the Enriquez connection, the monodromy relations of (\ref{2.3}) are invariant only under the subgroup $\mS_h \subset Sp(2h,\ZZ)$ that permutes a $\mB$-cycle into a $\mB$-cycle. Therefore, given a choice of canonical homology basis of cycles $\mA^I, \mB_I$, the degenerations on two different cycles $\mC$ and $\mC'$  are equivalent to one another only if $\mC$ and $\mC'$ are related by a permutation in $\mS_h$, which leaves an infinite number of equivalence classes.  In this letter, we shall examine only the example of pinching the cycle $\mA^h$, leaving the general case to subsequent work. In this particular situation, we shall find results that are strikingly intuitive and should guide us to organize the general case.

Fay's parametrization of a family of Riemann surfaces $\widehat \Sigma$ of genus $h$ near a non-separating divisor starts from a compact Riemann surface $\Sigma $ of genus $h-1$ with punctures $p_a,p_b$ and a complex parameter $q$ satisfying $|q| < 1$ \cite{Fay,DHoker:2017pvk}.\footnote{We depart from the notation of \cite{DHoker:2017pvk}, where the parameter $q$ of the non-separating degeneration and the blocks $\big ( \smallmatrix \Omega & v \cr v^t & \tau \endsmallmatrix \big )$ of the $h\times h$ period matrix in (\ref{3.3}) are denoted by $\mt$ and  $\big ( \smallmatrix \tau & v \cr v^t & \sigma \endsmallmatrix \big )$, respectively.} Local holomorphic coordinates, $z_a$ near $p_a$ and $z_b$ near $p_b$, are chosen such that $z_a(p_a)=0$ and $z_b(p_b)=0$. One  proceeds by drawing non-intersecting annuli $\cC_{a,b}= \{ z_{a,b}, \, |q|< |z_{a,b}| < 1 \}$  on $\Sigma$ with boundary circles $\mC_{a,b}'$ and $\mC_{a,b}''$, as shown in the figure below.

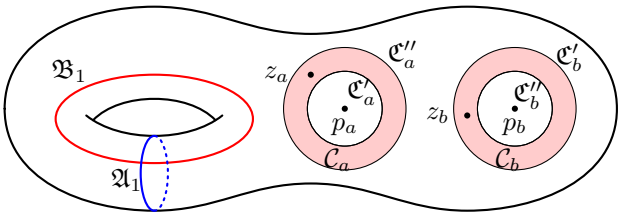
\sm 
\sm 
\begin{tikzpicture}[scale=0.9, line cap=round, line join=round]
\begin{scope}[xshift=-5cm,yshift=0cm]
    \draw[ thick] (0.5, 1.7) 
        to[out=90, in=180] (2.7, 3.2)
        to[out=0, in=180]  (5.0, 2.8)
        to[out=0, in=180]  (7.3, 3.2)
        to[out=0, in=90]   (9.5, 1.7)
        to[out=-90, in=0]  (7.3, 0.2)
        to[out=180, in=0]  (5.0, 0.6)
        to[out=180, in=0]  (2.7, 0.2)
        to[out=180, in=-90] (0.5, 1.7);
    \draw[ thick] (1.7, 1.6) .. controls (2.2, 1.2) and (3.2, 1.2) .. (3.7, 1.6);
    \draw[ thick] (1.8, 1.53) .. controls (2.2, 1.95) and (3.2, 1.95) .. (3.6, 1.53);
    %
    \draw[thick, red] (2.7, 1.55) ellipse (1.45 and 0.65);

    \draw[thick, dotted, blue] (2.7, 0.2) arc (-90:90:0.2 and 0.55);
        \draw[thick, blue] (2.7, 1.3) arc (90:270:0.2 and 0.55);
    \begin{scope}[shift={(5.5, 1.7)}]
        \draw[thin, fill=red!20] (0,0) circle (0.9);
        \draw[black, fill=white] (0,0) circle (0.55);
        \draw (0,0) circle (0.55);
        
        \fill[black] (0,0) circle (0.045);
        \node[below] at (0,0) {$p_a$};
        
        \fill[black] (-0.5, 0.5) circle (0.045);
        \node[left=2pt] at (-0.6, 0.5) {$z_a$};
        
        \node at (0.25, 0.25) {$\mC_a'$};
        \node at (0.85, 0.8)  {$\mC_a''$};
        \node[below=9pt] at (-0.1,-0.1) {$\cC_a$};
    \end{scope}
    \begin{scope}[shift={(8.0, 1.7)}]
        \draw[ thin, fill=red!20] (0,0) circle (0.9);
        \draw[black, fill=white] (0,0) circle (0.55);
        \draw (0,0) circle (0.55);
        
        \fill[black] (0,0) circle (0.045);
        \node[below] at (0,0) {$p_b$};
        
        \fill[black] (-0.7, -0.1) circle (0.045);
        \node[left=2pt] at (-0.75, -0.1) {$z_b$};
        
        \node at (0.2, 0.2)   {$\mC_b''$};
        \node at (0.8, 0.75)  {$\mC_b'$};
        \node[below=9pt] at (-0.1,-0.1) {$\cC_b$};
    \end{scope}
    \node at (2.3, 0.65) {$\mA_1$};
    \node at (1.45, 2.25) {$\mB_1$};

\end{scope}
\end{tikzpicture}

\sm

The discs enclosed by  $\mC_a'$ and $\mC_b''$ are removed and the annuli $\cC_a$ and $\cC_b$  are identified by the rule, 
\bea
\label{3.1}
z_a \, z_b = q
\eea
which identifies the curve  $\mC_a'$ with $\mC_b'$ and $\mC_a''$ with $\mC_b''$.   The punctures $p_a,p_b$ and the parameter $q$ provide the three extra local moduli of $\widehat\Sigma$ compared to those of $\Sigma$. 

A homology basis for $\widehat\Sigma$ is obtained from the cycles $\mA^\mu, \mB_\mu$  of $\Sigma$ for $\mu=1,\cdots, h{-}1$,  together with the cycle $\mA^h = - \mC_a'$ (the minus sign accounting for clockwise orientation of $\mA^h$) and the cycle $\mB_h$ between arbitrary points $\pi_a, \pi_b$ with coordinates $z_a, z_b$ on $\mathcal{C}_a$ and $\mathcal{C}_b$, respectively, such that $z_a z_b=q$. The dual basis of  holomorphic Abelian differentials on $\widehat\Sigma$ consists of the holomorphic differentials $\om_\mu$ of $\Sigma$ for $\mu=1,\cdots, h{-}1$, together with the meromorphic Abelian differential $\widehat \om_h$ defined 
through the combination of genus $h{-}1$ prime forms in (\ref{2.6a}) as
\begin{align}
 \widehat \om_h(x) = { \ki_{p_a  p_b}(x) \over 2 \pi i}  \,,   \qquad \! \! \! \oint _{\mA^\mu} \widehat \om_h=0   \,, \! \! \! \qquad  \oint _{\mA^h} \widehat\om_h=1 
    \label{3.2}
\end{align}
 The period matrix then takes the form,
\begin{equation}
\label{3.3}
   \widehat \Omega = \left ( \bma \Omega & v \cr v^t & \tau \ema \right ) \,, \qquad v_\mu = \oint _{\mB_\mu} \widehat\om_h = \int _\pa ^\pb \om_\mu
\end{equation}
while $\tau$ is given by,
\begin{equation}
    \label{3.4}
\tau = \int _{\pi_b}^{\pi_a} \! \widehat\om_h = { 1 \over 2 \pi i} \ln { E(\pi_a, p_a) E(\pi_b, p_b) \over E(\pi_a, p_b) E(\pi_b, p_a)}
= {\ln(q)  \over 2 \pi i}  + \cO(q^0)
\end{equation}
Therefore, up to terms that are exponential  in $q$, we have $q = e^{2 \pi i \tau}$ with $\Im(\tau) >0$. In particular, the prime form $\widehat E(x,y)$ on the surface $\widehat\Sigma$ of genus $h$ tends to the prime form $E(x,y)$ on $\Sigma$ of genus $h{-}1$,
\bea
\label{3.5}
\widehat E(x,y) = E(x,y) + \cO(q)
\eea
for fixed $x,y\in \Sigma$ outside the discs in the figure (see \cite{Fay} and especially the second paragraph of section 4.2 in~\cite{DHoker:2017pvk}).

\section{Degeneration of Enriquez kernels}
\label{sec:4}

The behaviour of Enriquez kernels under the above degeneration may be organized according to the \text{rank} $R=r{+}1$ of
$\widehat g^{I_1 \cdots I_r} {}_J(x,y)$ and its subset of indices equal to $h$. For this purpose, we use Greek letters $\mu,\nu,\cdots=1,\cdots,h{-}1$ for indices different from $h$ and capital Latin letters $I,J,\cdots$ for indices in the range $1,\cdots,h$, with the notation $\vec{\mu}$ and $\vec{I}$ for words in the respective indices. We shall first treat the cases of ranks 2 and 3 before proceeding to the case of arbitrary rank.

\sm

For rank 2, the representation (\ref{2.6})
of $\widehat g^I{}_J(x,y)$
implies,
\begin{align}
\!\!\widehat g^\mu{}_\nu (x,y) &  \to  g^\mu {}_\nu (x,y)\, , 
 \quad \quad  \
\widehat g^\mu {}_h (x,y)  \to  \frac{\chi_{p_a p_b}^\mu (x)}{2 \pi i}
\label{3.a.1} \\
\!\!\widehat g^h {}_\nu (x,y) & \to  0\, , 
 \quad  \quad  \quad
\widehat g^h {}_h (x,y)  \to  \frac{1}{2}\Big(  \chi _{p_a y} (x) {+} \chi_{p_b y} (x)\Big) 
\notag
\end{align}
up to terms of order $q$. 
The left column follows from the degeneration formula for the prime form in (\ref{3.5}) and the regularity of $\om_\nu$ near $p_a$ and $p_b$. For the top right formula, we spell out (\ref{2.6}) as follows, 
\begin{align}
\widehat g^\mu {}_h (x,y) & \rightarrow { 1 \over 2 \pi i} \int _{\mA^\mu} dt \,  \chi_{p_a  p_b}(t) \,  \chi_{t y}(x)   \notag \\
&=  { 1 \over 2 \pi i} \oint _{\mA^\mu} dt \, \chi_{p_a  p_b}(t)  \,  \chi(x,t) 
\label{3.a.2}
\end{align}
The contribution from the last term in (\ref{2.6}) vanishes  in view of $\delta^\mu_h=0$; 
the $y$-dependence in (\ref{3.a.2}) cancels in view of the first formula in (\ref{3.2}). Finally, we set $\vI=\emptyset$ in  (\ref{2.7a})  and use the remaining right side to re-express the second line of (\ref{3.a.2}) in the form given in (\ref{3.a.1}). The evaluation of $\widehat g^h{}_h (x,y)$ proceeds from (\ref{2.6}) as well, 
\begin{equation}
\label{3.a.3}
\widehat g^h {}_h (x,y) \rightarrow { 1 \over 2 \pi i} \oint _{\mA^h} dt \,
 \chi_{p_a p_b}(t) \, \chi_{t y}(x) 
- \half \, \chi_{p_a  p_b}(x) 
\end{equation}
Representing $\mA^h$ by the cycle $- \mC_a'$, the integral can be evaluated from its residue at $p_a$,
\bea
\label{3.a.4}
\widehat g^h {}_h (x,y) \rightarrow  \chi_{p_a y}(x) - \half \, \chi_{p_a  p_b}(x)
\eea
which by (\ref{2.6a}) gives the bottom right formula of~(\ref{3.a.1}). 

\sm

For rank 3, the degenerations of the components for the Enriquez kernel $\widehat g^{KI}{}_J(x,y)$ may be obtained by the same processes as was used for rank 2, and we find,
\begin{align}
\label{3.a.5}
\widehat g ^{\mu \nu}{}_{\lambda}(x,y)
&\rightarrow \, g^{\mu \nu}{}_{\lambda}(x,y) \, ,  \qquad
\widehat g^{\mu \nu}{}_h(x,y)
\rightarrow  \, 
\frac{ \chi^{\mu \nu} _{p_a p_b} (x) }{2\pi i}
\no \\
\widehat g^{\mu h }{}_{\nu}(x,y)
&\rightarrow \, 0 \, , \quad \ \
\widehat g^{\mu h }{}_h (x,y)
\rightarrow \, \frac{1}{2}\Big(  \chi^\mu _{p_a y} (x) {+} \chi^\mu_{p_b y} (x)\Big) 
\notag \\
\widehat g ^{h \mu }{}_{\nu}(x,y)
&\rightarrow \, 0\, , \quad \ \
\widehat g ^{h \mu }{}_h (x,y)
\rightarrow \, 0 \notag \\
\widehat g ^{hh }{}_{\nu}(x,y)
&\rightarrow \, 0 \, , \quad \ \
\widehat g ^{hh}{}_h(x,y)
\rightarrow \, \frac{i\pi}{6} \,  \chi_{p_a p_b} (x) 
\end{align}
We shall provide the derivation for the expression in the bottom right of (\ref{3.a.5}). The starting point is again (\ref{2.7}) evaluated now for $K=\vI=J=h$,
\begin{align}
    &\widehat g ^{hh}{}_h(x,y) -  { 2 \pi^2 \over 3} \widehat \om_h(x) = - \oint _{\mA^h} dt \, \widehat g^L{}_h(x,t) \, \widehat g^h{}_L(t,y)  \notag\\
    &  \to   \frac{1}{4} \oint _{\mA^h}\!\!\! dt \,  \Big(  \chi_{p_a t} (x) {+} \chi_{p_b t} (x)\Big) \Big(  \chi_{p_a y} (t) {+} \chi_{p_b y} (t)\Big) \! \! \! \label{rk3ex} 
\end{align}
where in the second line we used the fact that the contribution from $1 \leq L\leq  h{-}1$ cancels by the bottom left relation in (\ref{3.a.1}). 
The subtraction on the left side of (\ref{rk3ex}) degenerates to ${ 2 \pi^2 \over 3} \widehat\om_h(x) \rightarrow -{ i \pi \over 3} \chi_{p_a p_b} (x)  $, and the integral in the second line evaluates to ${ i \pi \over 2} \chi_{p_a p_b} (x)$ via~residues.\footnote{Recall that the residue of $\chi_{yz}(t)$ at $y$ is $-1$ and at $z$ is $+1$.}

\sm

For arbitrary rank $\geq 2$, the myriad cases, whether the indices take the value $h$ or not, follow a strikingly simple pattern which is a first main result of this work, 
\begin{subequations}
\label{3.b.1}
\begin{align}
 \widehat g^{\vmu}{}_\nu (x,y) & \to \, g^{ \vmu}{}_\nu (x,y) 
\label{3.b.1aa}
\\
\widehat g^{\cdots h \cdots }{}_\nu  (x,y) & \to \, 0 
\label{3.b.1bb} \\
\label{3.b.1cc}
\widehat g^{\cdots h \mu \cdots }{}_h (x,y) & \to \, 0 
\\
\widehat g^{\vmu h }{}_h (x,y)
& \to \, \frac{1}{2}\Big(  \chi^\vmu _{p_a y} (x) + \chi^\vmu_{p_b y} (x)\Big)
\label{3.b.1dd} 
\\
\widehat g^{\vmu \, \vh_\ell }{}_h (x,y)&  \to \, {(2 \pi i)^{\ell-1} \over \ell! } \, {\Ber_\ell }\,   \chi_{p_a p_b}^\vmu (x)
   \quad \ \ \ell \neq 1
\label{3.b.1ee}
\end{align}
\end{subequations}
where $\vh_\ell$ is the word consisting of $\ell$ repetitions of the letter $h$ and no other letters.\footnote{Note that, for odd $\ell\geq 3$, the Bernoulli number $\Ber_\ell$ and therefore the non-separating degeneration of $\widehat g^{\vmu \, \vh_\ell }{}_h (x,y)$ vanishes.} One easily verifies that the different arrangements of indices cover all possible cases for the decomposition of $\widehat g^\vI{}_J(x,y)$: the first two cover the cases where $J \not= h$ while the next three cover the cases when $J=h$. The notation in (\ref{3.b.1bb}) and (\ref{3.b.1cc}) indicates that kernels $\widehat g^{\cdots h \cdots }{}_\nu$ with at least one upper index $h$ and $\widehat g^{\cdots h \mu \cdots }{}_h$ with a sequence $h\mu$ among the upper indices have vanishing degenerations, also if one or both of the ellipses $\cdots$ are empty. Given that the right sides of all cases are expressible in terms of $g^{\vec{\mu}}{}_\nu$ at genus $h{-}1$, 
possibly projected to their trace components $\chi$ depending on $p_a,p_b$, we conclude that
Enriquez kernels close under pinching of the cycle $\mA^h$. The proof of the relations is presented in the appendix.

Note that, while the first two relations (\ref{3.b.1aa}), (\ref{3.b.1bb}) can be found in section 7 of \cite{Baune:2025sfy} for those regions in moduli space where the underlying Poincar\'e series converge, the remaining degeneration formulas of (\ref{3.b.1}) are~new.

The following corollary of (\ref{3.b.1}) for the trace components in (\ref{2.5}) and (\ref{chifun}) will be useful for the next section,
\begin{align}
\label{5.1}
\widehat \chi ^\vmu_{yz} (x) & \, \to \,  \chi ^\vmu_{yz} (x) 
 \, , \qquad \quad
\widehat \chi ^{\cdots h \cdots}_{yz} (x)
  \, \to \, 0 
\end{align}
with vanishing degeneration in the second cases whenever one or more of the upper indices is $h$.

\section{Degeneration of Enriquez connection}
\label{sec:5}
We shall now deduce the degeneration of the Enriquez connection in \eqref{2.4} from the degenerations (\ref{3.b.1}) and (\ref{5.1}) of its components $g$ and $\chi$. With the shorthand
\begin{align}
     \widehat{\bf{W}}_{J}(x_i;B_i) & =     \sum _{\vec{I}} \widehat\varpi ^{\vec{I}} {}_{J} (x_i) \, B_{i \vec{I}} \,,\notag \\
  \widehat{\bf{X}} (x_i,x_j ;B_i) & =   \sum _{\vec{I}} \widehat \chi ^{ \vec{I}} (x_i,x_j)\, B_{i \vec{I}}
  \label{WXdef}
\end{align}
for the generating series of  (1,0) forms in $x_i$ with $B_{iI} = {\rm ad}_{b_{iI}}$, the
Enriquez connection of (\ref{2.4}) on the universal cover of the genus-$h$ surface $  \widehat \Sigma $ takes the form 
\begin{align}
\widehat \cK_\text{E} (\xx) &=  \sum_{i=1}^n dx_i \, \widehat K_i(\xx)  \label{KEgh} \\
\widehat K_i(\xx) &=   \widehat {\bf{W}}_J(x_i;B_i) \, a^J_i + \sum_{j\neq i}^n \widehat {\bf{X}}(x_i,x_j;B_i) \, t_{ij} 
\notag
\end{align}
The analogous $(1,0)$-form components $K_i(\xx)$ in $x_i$ of the Enriquez connection $\cK_\text{E} (\xx) $ at genus $h{-}1$ are given by 
\begin{equation}
 K_i(\xx) = {\bf{W}}_\mu(x_i;B_i)\,  a^\mu_i + \sum_{j\not=i}^n {\bf{X}}(x_i,x_j;B_i) \, t_{ij} 
    \label{5.3}
\end{equation}
in terms of genus $h{-}1$ instances of the generating series \eqref{WXdef} with a sum over words $\vec\mu = \mu_1\cdots \mu_r$ instead of $\vI$ whose letters $\mu_i$ are again restricted to $1, \cdots, h{-}1$.
The non-separating degenerations (\ref{3.b.1}) and (\ref{5.1}) of the kernels can be readily translated into those of the connection (\ref{KEgh}) by  eliminating each $\widehat\varpi(x)$ in favor of $\widehat \chi(x,y)$ and $\widehat g(x,y)$ via (\ref{2.5}) (involving an arbitrary second point $y$ on the cover of $\widehat\Sigma$) and using the last structure relation of $\mt_{h,n}$ in (\ref{2.2}). The result can be presented in unified form
\begin{align} 
&\widehat K_i(\xx) \rightarrow K_i(\xx) + \sum_{\ell=0}^\infty { (2 \pi i )^{\ell-1} \over \ell !} 
 \label{5.4}\\
& \ \  \times 
\Big ( \Ber_\ell ^+ {\bf{X}}  (x_i, p_a;B_i)  - \Ber_\ell ^- {\bf{X}} (x_i,p_b;B_i) \Big )  (B_{i  h})^\ell a_i^h
\notag
\end{align}
by introducing the variants $\Ber_\ell^\pm$ of the Bernoulli numbers with  $\Ber_\ell^\pm = {\Ber_\ell }$ for $\ell \neq 1$ and $\Ber_1^\pm=\pm \frac{1}{2}$. 
Using the generating functions for the Bernoulli numbers,
\bea
\label{5.5}
{ \pm x \over e^{ \pm 2\pi i x} -1} = \sum _{\ell=0}^\infty  (2 \pi i )^{\ell-1} { \Ber ^\mp _\ell \over \ell !} \, x^\ell
\eea
the degeneration of the connection components $\widehat K_i$ can be brought into the intuitive form, 
\beq
\widehat K_i(\xx) \rightarrow K_i(\xx) \! + \!
{\bf{X}} (x_i, p_a;B_i)   t_{ia}  \! + \! {\bf{X}}  (x_i, p_b;B_i)   t_{ib} 
\label{5.6} 
\eeq
where the composite generators $t_{ia}$, $t_{ib}$ are defined as,
\begin{align}
    \label{5.7}
t_{ia} =  { B_{ih} \over 1- e^{- 2 \pi i B_{ih}} } a^h_i \,, \qquad 
t_{ib} =  { B_{ih} \over 1- e^{2 \pi i B_{ih}} } a^h_i 
\end{align}
Equation \eqref{5.6} is one of the main results of this letter,
stating that the degeneration of a genus $h$ Enriquez connection is given by an Enriquez connection with $n$ variables on a Riemann surface of genus $h{-}1$ in the presence of two punctures  $p_a$ and $p_b$. The residues at these punctures are composite generators (\ref{5.7}) of $\mt_{h,n}$ given by Lie series in the last components $a_i^h$ and $b_{ih}$, and the notation
$t_{ia}$, $t_{ib}$ aims to express their analogy with the $t_{ij}$ generators along with the poles $(x_i{-}x_j)^{-1}$ of (\ref{2.4}).
The characteristic series (\ref{5.7}) in Bernoulli numbers already appeared in the degeneration of the Calaque-Enriquez-Etingof (CEE) connection \cite{CEE} at genus one and are a general hallmark of two nodal points obtained from a pinched homology cycle.

\bigskip

\section{Sequential degenerations}

A simple corollary of the above results is the sequential degeneration of the multi-variable Enriquez connection which eventually yields a Knizhnik-Zamolodchikov (KZ) connection on a genus 0 surface with $2h$ additional nodal points, as depicted below for the genus 3 case.

\begin{widetext}
 \begin{figure} [h]

\begin{tikzpicture}[xscale=.58,yscale=.58, line cap=round, line join=round]

    \begin{scope}[xshift=0cm, yshift=0cm]
        \draw[thick] (-5.8,0) to[out=90,in=180] (-4.2,1.4) 
                             to[out=0,in=180] (-2.8,0.9)
                             to[out=0,in=180] (-1.4,1.4)
                             to[out=0,in=180] (0,0.9)
                             to[out=0,in=180] (1.4,1.4)
                             to[out=0,in=90] (2.8,0)
                             to[out=-90,in=0] (1.4,-1.4)
                             to[out=180,in=0] (0,-0.9)
                             to[out=180,in=0] (-1.4,-1.4)
                             to[out=180,in=0] (-2.8,-0.9)
                             to[out=180,in=0] (-4.2,-1.4)
                             to[out=180,in=-90] (-5.8,0);

        \begin{scope}[xshift=-5pt]
                    \draw[thick] (-5.0, 0.1) .. controls (-4.5, -0.25) and (-3.7, -0.25) .. (-3.1, 0.1);
        \draw[thick] (-4.8, 0.0) .. controls (-4.4, 0.25)  and (-3.8, 0.25)  .. (-3.4, 0.0);
        \end{scope}

        \draw[thick] (-2.3, 0.1) .. controls (-1.8, -0.25) and (-1.0, -0.25) .. (-0.5, 0.1);
        \draw[thick] (-2.1, 0.0) .. controls (-1.7, 0.25)  and (-1.1, 0.25)  .. (-0.7, 0.0);

        \draw[thick] (0.5, 0.1) .. controls (1.0, -0.25) and (1.8, -0.25) .. (2.3, 0.1);
        \draw[thick] (0.7, 0.0) .. controls (1.1, 0.25)  and (1.7, 0.25)  .. (2.1, 0.0);
    \end{scope}

    \draw[->, ultra thick] (3.3, 0) -- (4.3, 0);

    \begin{scope}[xshift=9.4cm, yshift=0cm]
        \draw[thick] (-4.5,0) to[out=90,in=180] (-2.8,1.4) 
                             to[out=0,in=180] (-1.4,0.9)
                             to[out=0,in=180] (0,1.4)
                             to[out=0,in=90] (1.4,0)   
                             to[out=-90,in=0] (0,-1.4)
                             to[out=180,in=0] (-1.4,-0.9)
                             to[out=180,in=0] (-2.8,-1.4)
                             to[out=180,in=-90] (-4.5,0);

    \begin{scope}[xshift=-5pt]
        \draw[thick] (-3.6, 0.1) .. controls (-3.1, -0.25) and (-2.3, -0.25) .. (-1.8, 0.1);
        \draw[thick] (-3.4, 0.0) .. controls (-3.0, 0.25)  and (-2.4, 0.25)  .. (-2.0, 0.0);
 \end{scope}

        \draw[thick] (-0.9, 0.1) .. controls (-0.4, -0.25) and (0.4, -0.25)  .. (0.9, 0.1);
        \draw[thick] (-0.7, 0.0) .. controls (-0.3, 0.25)  and (0.3, 0.25)   .. (0.7, 0.0);

        \coordinate (pa1) at (1, 0.45);
        \coordinate (pb1) at (1, -0.45);
        \fill[blue] (pa1) circle (2pt) node[left, black, xshift=2pt] {$p_{a_3}$};
        \fill[blue] (pb1) circle (2pt) node[left, black, xshift=2pt] {$p_{b_3}$};
        \draw[blue, thick, dashed] (pa1) .. controls (2.2, 1.2) and (2.2, -1.2) .. (pb1);
    \end{scope}

    \draw[->, ultra thick] (11.8, 0) -- (12.8, 0);

    \begin{scope}[xshift=16.2cm, yshift=0cm]
        \draw[thick] (-2.4,0) to[out=90,in=180] (-0.8,1.4)
                             to[out=0,in=90] (0.8,0)   
                             to[out=-90,in=0] (-0.8,-1.4)
                             to[out=180,in=-90] (-2.4,0);

        \draw[thick] (-1.5, 0.1) .. controls (-1.1, -0.25) and (-0.5, -0.25) .. (-0.1, 0.1);
        \draw[thick] (-1.3, 0.0) .. controls (-1.0, 0.25)  and (-0.6, 0.25)  .. (-0.3, 0.0);

        \coordinate (m_pa1) at (0.4, 0.4);
        \coordinate (m_pb1) at (0.4, -0.4);
        \fill[blue] (m_pa1) circle (2pt) node[left, black, xshift=2pt] {$p_{a_3}$};
        \fill[blue] (m_pb1) circle (2pt) node[left, black, xshift=2pt] {$p_{b_3}$};
        \draw[blue, thick, dashed] (m_pa1) .. controls (1.5, 1.1) and (1.5, -1.1) .. (m_pb1);

        \coordinate (m_pa2) at (-2.0, 0.4);
        \coordinate (m_pb2) at (-2.0, -0.4);
        \fill[blue] (m_pa2) circle (2pt) node[right, black, xshift=-2pt] {$p_{a_2}$};
        \fill[blue] (m_pb2) circle (2pt) node[right, black, xshift=-2pt] {$p_{b_2}$};
        \draw[blue, thick, dashed] (m_pa2) .. controls (-3.1, 1.1) and (-3.1, -1.1) .. (m_pb2);
    \end{scope}

   
    \draw[->, ultra thick] (18, 0) -- (19, 0) node[midway, above=4pt] {};

    \begin{scope}[xshift=22cm, yshift=0cm]

        \draw[gray, thick, dashed] (1.5,0) arc (0:180:1.5cm and 0.4cm);

        \draw[thick] (0,0) circle (1.5cm);

        \draw[gray, thick] (-1.5,0) arc (180:360:1.5cm and 0.4cm);

        \coordinate (pa1) at (1.1, 0.45);
        \coordinate (pb1) at (1.1, -0.45);

        \fill[blue] (pa1) circle (2pt) node[right, black, xshift=1pt, yshift=10pt] {$p_{a_3}$};
        \fill[blue] (pb1) circle (2pt) node[right, black, xshift=1pt, yshift=-10pt] {$p_{b_3}$};

        \draw[blue, thick, dashed] (pa1) .. controls (2.2, 1.3) and (2.2, -1.3) .. (pb1);

        \coordinate (pah) at (-1.1, 0.45);
        \coordinate (pbh) at (-1.1, -0.45);

        \fill[blue] (pah) circle (2pt) node[above, black, xshift=-10pt, yshift=2.4pt] {$p_{a_2}$};
        \fill[blue] (pbh) circle (2pt) node[below, black, xshift=-10pt, yshift=-2pt] {$p_{b_2}$};

        \draw[blue, thick, dashed] (pah) .. controls (-2.2, 1.3) and (-2.2, -1.3) .. (pbh);

        \coordinate (pa2) at (-0.2, 1);
        \coordinate (pb2) at (-0.2, -1);
        
        \fill[blue] (pa2) circle (2pt) node[right, black, xshift=-1.pt, yshift=2pt] {$p_{a_1}$};
        \fill[blue] (pb2) circle (2pt) node[right, black, xshift=-1pt, yshift=-2pt] {$p_{b_1}$};

        \draw[blue, thick, dashed] (pa2) .. controls (.3,0) and (.3,0) .. (pb2);

    \end{scope}

\end{tikzpicture}
\end{figure}
\end{widetext}
By pinching all the $\mA^I$ cycles except for $\mA^1$, a simple iteration of \eqref{5.6} (using (\ref{5.1}) and (\ref{2.2}) to accommodate for the additional punctures) yields,
\begin{align}
&\widehat K_i(\xx) \xrightarrow{(h-1)} a_i^1 + \sum^n_{j\neq i}  {\bf{X}}_1  (x_i, x_j;B_{i1}) \, t_{ij}
\label{E_deg_hm1} \\
&\quad + \sum _{\lambda=2}^h
 \Big( 
  {\bf{X}}_1  (x_i, p_{a_\lambda};B_{i1}) \, t_{ia_\lambda}
  + {\bf{X}}_1  (x_i, p_{b_\lambda};B_{i1}) \, t_{ib_\lambda}
  \Big)
 \notag
\end{align}
where $p_{a_\lambda}$, $p_{b_\lambda}$ with $\lambda = 2,\cdots,h$ are $2h{-}2$ nodal points on the torus. The associated residues of the poles in the connection (\ref{E_deg_hm1}) are the same characteristic Bernoulli series as seen in (\ref{5.7}):
\beq
t_{ia_\lambda} =  \frac{B_{i\lambda}}{1 - e^{-2\pi i B_{i\lambda}}} \,a_i^{\lambda} \, , \ \ \ \ \ \
t_{ib_\lambda} =  \frac{B_{i\lambda}}{1 - e^{2\pi i B_{i\lambda}}} \, a_i^{\lambda}
\label{gents}
\eeq 
The simple form of (\ref{E_deg_hm1}) comes from the fact that the components $\varpi^\vI{}_J(x)$ of the series ${\bf W}_J$ in (\ref{WXdef}) vanish at genus one if $\vI \neq \emptyset$ and otherwise reduce to $\varpi^\emptyset{}_1(x)=~1$. All the dependence of (\ref{E_deg_hm1}) on the points is carried by the genus-one instance ${\bf X}_1$ of the series (\ref{WXdef}) in $\chi ^{ \vec{I}} (x_i,x_j)$ which reduce to Kronecker-Eisenstein kernels $g^{(r)}(z)$ at~$r\geq 1$, 
\begin{equation}
g^{I_1 \cdots I_r}{}_J(x,y) \big|_{h=1}
=  g^{(r)}(x{-}y) =
 - \chi^{\vec{1}_{r-1}}(x,y)  \big|_{h=1}  
\end{equation}
with generating series
\begin{align}
{\bf X}_1(x,y;B) &= - \sum^\infty_{r=1} g^{(r)}(x{-}y) \, B^{r-1}
\label{xinKE} \\
&= \frac{1}{B} - \frac{\theta_1'(0) \theta_1(x{-}y{+}B)}{\theta_1(x{-}y) \theta_1(B)} \notag
\end{align}
On these grounds, the sequential degeneration (\ref{E_deg_hm1}) of the Enriquez connection to genus $h=1$ can be identified with the CEE connection on the configuration space of $n$ points on the universal cover of the torus \cite{CEE} with $2h{-}2$ additional punctures $p_{a_\lambda}$, $p_{b_\lambda}$.

As a last step, we degenerate to genus 0 by pinching the left-over cycle $\mA^1$. The degeneration of the genus one connection is best expressed in terms of coordinates on the nodal sphere defined via  $\sigma_i = e^{2\pi i x_i}$ as well as $\sigma_{p_{a_\lambda}} = e^{2\pi i p_{a_\lambda}}$ and $\sigma_{p_{b_\lambda}} = e^{2\pi i p_{b_\lambda}}$. Based on the degeneration of the Kronecker-Eisenstein series,
\begin{equation}
dx_i\, \frac{\theta_1'(0) \theta_1(x_{i}{-}x_j{+}B_{i1})}{\theta_1(x_{i}{-}x_j) \theta_1(B_{i1})}\, B_{i1}  a_i^1 
\rightarrow d \sigma_i \, \bigg( 
\frac{B_{i1} a_i^1}{\sigma_i{-}\sigma_j} -\frac{t_{ib_1}}{\sigma_i}  \bigg)
\label{h01sec1}
\end{equation}
with $t_{ib_1}$ in (\ref{gents}), and the $\mt_{h,n}$ relation due to  (\ref{2.2})
\bea
B_{i1} a_i^1 = - \sum_{\lambda=2}^h ( t_{ia_\lambda}+ t_{ib_\lambda} ) - \sum^n_{j\neq i} t_{ij}
\label{h01sec2}
\eea
we can express the sequential degeneration of the Enriquez connection where all the $\mA$-cycles are pinched as  
\begin{align}
dx_i \, &\widehat K_i(\xx)  \xrightarrow{(h)}
  -  \sum^n_{j\neq i} \frac{d\sigma_i}{\sigma_i{-}\sigma_j} t_{ij}  \label{h0degen} \\
&-  \sum_{\lambda=1}^{h} 
\bigg(
\frac{d \sigma_i}{\sigma_i{-}\sigma_{a_\lambda}}
\, t_{ia_\lambda}
{+} \frac{d \sigma_i}{\sigma_i{-}\sigma_{b_\lambda} }
\,  t_{ib_\lambda} 
\bigg) 
\notag
\end{align}
The terms with $\lambda = 1$ in the last line were lined up with the contributions from the remaining nodal points at $2\leq \lambda \leq h$ by introducing the nodal points $\sigma_{b_1} = 0$ and $\sigma_{a_1} \rightarrow \infty$. Their residues $t_{ib_1}$ and $t_{ia_1}$ follow from (\ref{h01sec1}) and (\ref{h01sec2}) with $B_{i1} a_i^1 = t_{ib_1}+t_{ia_1}$, respectively.

Hence, the degeneration (\ref{h0degen}) of the $n$-variable Enriquez connection where all the $\mA$-cycles are pinched reproduces a KZ connection on the Riemann sphere with $n$ variables $\sigma_i$ and $2h$ additional punctures $\sigma_{a_\lambda}$, $\sigma_{b_\lambda}$. The residues of the poles $ t_{ib_\lambda} $ and $ t_{ia_\lambda} $ as $\sigma_i$ approaches the $2h$ nodal points generalize the well-known generating series (\ref{gents}) of Bernoulli numbers in the degeneration limit of the CEE connection on the nodal sphere \cite{CEE}.

\section{Degeneration of DHS kernels}

While this letter has mainly focused on the degeneration of the meromorphic and multivalued Enriquez kernels and connection, we would like to briefly discuss the analogous degeneration of the DHS kernels introduced in \cite{DHoker:2023vax}. Unlike the Enriquez connection, the DHS connection of \cite{DHoker:2026lgg} is single valued and non-meromorphic on the configuration space of $n$ points on a genus-$h$ surface. The modular invariance of the DHS connection introduces two types of falloff behaviour into its asymptotics in the
non-separating degeneration which are governed by the \emph{real degeneration parameter} $t$ \cite{DHoker:2017pvk}
\begin{equation}
    t= \frac{\det (\Im \widehat \Omega)}{\det (\Im \Omega)} 
    \label{deft}
\end{equation}
where $\widehat \Omega$ and $\Omega$ denote the $h\times h$ and $(h{-}1)\times (h{-}1)$ period matrices (\ref{2.1}) on $\widehat \Sigma$ and $\Sigma$, respectively. Apart from an infinity of exponentially suppressed terms in the degeneration $t \rightarrow \infty$, DHS kernels exhibit polynomials in $1/t$ as their leading behaviour in this limit, similar to the asymptotics of higher-genus modular graph forms in \cite{DHoker:2017pvk, DHoker:2020tcq} but excluding positive powers of~$t$.

In the remainder of this section, we shall spell out the leading $t^0$ term of these  polynomials in the non-separating degenerations of arbitrary DHS kernels $f^{I_1\cdots I_r}{}_J$ and leave the investigation of the $1/t$ suppressed contributions beyond the simplest case of $r=1$ for later work. As we will see, the leading $t \rightarrow \infty$ asymptotics of DHS kernels exhibits striking structural parallels to the non-separating degenerations (\ref{3.b.1}) of Enriquez kernels in the same way as their Fay identities \cite{DHoker:2024ozn, Baune:2024ber, DHoker:2026ggx} and moduli variations \cite{DHoker:2025dhv, DESZ:2026} can be formally mapped into one another via substitutions $f^{\vI}{}_J \leftrightsquigarrow g^{\vI}{}_J$.

While Enriquez kernels are built from $\mA$-cycle convolutions of the meromorphic prime form, DHS kernels are recursively constructed from surface integrals of the single-valued and modular invariant Arakelov Green function $\mathcal{G}(x,y)$ \cite{Faltings:1984, Alvarez-Gaume:1986nqf,  DHoker:2017pvk}. We shall omit the hat notation for the surface $\Sigma$ of genus $h$ and associated quantities in the subsequent equations (\ref{tmp.01}) to (\ref{diffcgmult}) to avoid cluttering.

\sm

The simplest Enriquez kernels (\ref{2.6}) closely follow the structure of the rank-two DHS kernels
\begin{equation}
f^I{}_J(x,y) = \int_{\Sigma} d^2 z \, \bar \omega^I(z) \, \omega_J(z)\, \p_x \cG_{zy}(x)
\label{tmp.01}
\end{equation}
involving antiholomorphic Abelian differentials
\bea
\bar \omega^I(x) = Y^{IJ} \bar \omega_J(x) \, , \ \ \ \ \ \ Y^{IJ} = \big[(\Im \Omega)^{-1}\big]^{IJ}
\label{ombar}
\eea
and differences of Arakelov Green functions 
\bea
\cG_{yz}(x) = \cG(x,y) - \cG(x,z)
\label{diffcg}
\eea
Higher-rank DHS kernels are obtained from repeated convolutions ($\vI \neq \emptyset$),
\begin{align}
f^{K\vec{I}}{}_J(x,y) &= \int_{\Sigma} d^2 z \, \partial_x \cG(x,z)\, \bar \omega^K(z) \,f^{\vec{I}}{}_J(z,y) 
\end{align}
and similar to (\ref{2.5}) decompose into $y$-independent traceless parts $\partial_x \Phi^{\vec{I}K}{}_J(x)$ and Green
 functions in the trace,
\beq
f^{\vec{I}K}{}_J(x,y)  = \partial_x \Phi^{\vec{I}K}{}_J(x)- \partial_x \cG^{\vec{I}}(x,y)\, \delta^{K}_J 
\eeq
The notation exposes that DHS kernels are derivatives of single-valued functions on one or two copies of $\Sigma$, namely
\beq
\cG^{K\vec{I}}(x,y) = \int_{\Sigma} d^2 z \,  \cG(x,z)\, \bar \omega^K(z) \,\p_z \cG^{\vI}(z,y)
\label{defallg}
\eeq
(with $\cG^{\emptyset}(x,y) = \cG(x,y)$) and traceless tensors $\Phi^{\vec{I}K}{}_J(x)$ obtained from similar iterative convolutions of
\beq
\Phi^I{}_J(x) = \int_{\Sigma} d^2 z \,\cG(x,z) \, \bar \omega^I(z) \,\omega_J(z)
\label{defphit}
\eeq
We will make frequent use of the shorthand
\bea
\cG^{\vI}_{yz}(x) = \cG^{\vI}(x,y) - \cG^{\vI}(x,z)
\label{diffcgmult}
\eea
which generalizes (\ref{diffcg}) to non-empty $\vI$ and resembles $\chi^\vI_{yz}(x)$ for Enriquez kernels in (\ref{chifun}).

\sm

In the subsequent discussion of non-separating degenerations, we again denote quantities at genus $h$ and $h{-}1$ with and without a hat, respectively. For the inverse imaginary part (\ref{ombar}) of the period matrix, we have the block decomposition \cite{DHoker:2017pvk} (with $(Y{\cdot} \Im v)^\mu = Y^{\mu \lambda} \Im v_\lambda$)
\begin{equation}
    \widehat Y^{IJ} =  \left( \begin{matrix} Y^{\mu \nu} &\! 0 \\ 0 &\! 0 \end{matrix} \right) + \frac{1}{t}  \left( \begin{matrix} (Y{\cdot} \Im v)^\mu (Y{\cdot} \Im v)^\nu &- (Y{\cdot} \Im v)^\mu \\ - (Y{\cdot} \Im v)^\nu &1 \end{matrix} \right)
\label{mhh.10}
\end{equation}
featuring the degeneration parameter $t$ in (\ref{deft}) and the $\mB$ periods $v_\mu$ in (\ref{3.3}). Moreover, following \cite{DHoker:2017pvk}, we shall perform a non-holomorphic change of basis with respect to the indices $I,J,\cdots = 1,\cdots,h$ 
to attain single-valued functions of the nodal points $p_a,p_b$ in the individual components. This is most easily illustrated through the identification of the Abelian differentials  
\begin{align}
   &   \widehat \omega_\mu =  \omega_\mu \,, \qquad  \widehat \omega_h(x) =  \frac{1}{2\pi i } \, \p_x \ln {E(x,p_a) \over E(x,p_b)} \notag \\ 
     &\overline{\widehat{\omega}}^\mu =   \bar  \omega^\mu  + \frac{1}{t}\, ( Y{\cdot} \Im v )^{\mu} \, \Big(  ( Y{\cdot} \Im v )^{\nu} \bar\omega_\nu- \bar \omega_h \Big)  \notag \\ 
         &\overline{\widehat{\omega}}^h =   - \frac{1}{t}\, \Big(  ( Y{\cdot} \Im v )^{\nu} \bar\omega_\nu- \bar \omega_h \Big)  
         \label{degab}
\end{align}
The $\mB$-cycle monodromies in $p_a$, $p_b$ of the prime forms and $\Im v$ on the right sides can be compensated by trading the basis elements $\widehat\omega_h(x)$ and $\overline{\widehat{\omega}}^\mu(x) $ for
\begin{align}
\widehat\omega_t(x) &= \widehat\omega_h(x)-\widehat\omega_\mu(x) \, ( Y{\cdot} \Im v )^{\mu}
\label{mutbasis}\\
\overline{\widehat{\omega}}^{ \mu_t}(x) &=  \overline{\widehat{\omega}}^\mu(x) +  ( Y{\cdot} \Im v )^{\mu}\,   \overline{\widehat{\omega}}^h(x) \notag
\end{align}
while retaining $\widehat \omega_\mu$ and
$\widehat {\bar \omega}^h$ in the basis. This choice results in the following equivalents to (\ref{degab}) which are term by term single-valued in $p_a,p_b$, 
\begin{align}
      \widehat \omega_\mu (x) &=  \omega_\mu (x) \,, &  \widehat \omega_t (x) &=  \frac{1}{2\pi i} \,\p_x \cG_{p_a p_b}(x) \label{nwomdeg}\\ 
    \overline{\widehat{\omega}}^{ \mu_t}(x)  &=    \bar \omega^\mu (x) \,,&\overline{\widehat{\omega}}^{h}(x)  &=   - \frac{1}{2\pi i t} \, \p_{\bar x}\cG_{p_a p_b}(x)   \notag  
\end{align}
and rely on the link between
the prime form and the Arakelov Green function,
\beq
\p_x \log \frac{E(x,y)}{E(x,z)} = \p_x \cG_{zy}(x)  + 2\pi i \, \omega_\mu(x) \,\Im \int^y_z  \omega^\mu
\label{mhh.09}
\eeq
In contrast to (\ref{3.5}) for the prime form, the non-separating degeneration of the Arakelov Green function introduces a $1/t$ correction besides exponentially suppressed terms,
\begin{align}
\p_x\widehat{\cG}_{yz}(x)\rightarrow \p_x \cG_{yz}(x)+ \frac{\p_x\cG_{p_a p_b}(x) }{4\pi t} \,\big( \mgh_{p_a p_b}(z){-}\mgh_{p_a p_b}(y) \big) 
\end{align}
With these prerequisites, one can show that the rank two DHS kernels (\ref{tmp.01}) enjoy the non-separating degenerations,
\begin{align}
 &  \! \! \hat{f}^{ \mu_t}\,_{\nu}(x,y) \to   f^{{\mu}}{}_\nu (x,y) \label{degfrk2} \\
 &\qquad \qquad \quad - \frac{ 1}{4\pi t} \,\p_x \cG_{p_a p_b} (x) \,\Big(  \Phi_{p_a p_b} {}^{\! \mu} {}_\nu{-}\delta^\mu_\nu    \mgh_{p_a p_b}(y) \Big)  \notag \\
&\! \! \hat{f}^{ \mu_t}\,_{t}(x,y) \to\frac{1}{2\pi i } \, \p_x\cG_{p_a p_b}^\mu(x) \notag \\
    &\!  \! \hat{f}^{h}\,_{\nu}(x,y) \to  \frac{i}{2 t} \, \omega_\alpha(x) \Big(\Phi_{p_a p_b}{}^{\alpha} {}_\nu - \delta^\alpha_\nu  \mgh_{p_a p_b}(x)   \Big)  \notag \\
 &   \hat{f}^{h}\,_{t}(x,y) \to \frac{1}{2}\,\Big( \p_x \cG_{p_a y}(x)+\p_x \cG_{p_b y}(x) \Big)  \notag \\
 & \qquad \qquad \quad + \frac{1}{4\pi t}\, \p_x \cG_{p_a p_b}(x) \, \Big( \mgh_{p_a p_b}(y)-\mgh_{p_a p_b}(x)\Big)  \notag
 \end{align}
using the shorthand $\Phi_{yz}{}^\mu{}_\nu =
\Phi^\mu{}_\nu(y) - \Phi^\mu{}_\nu(z)$ for differences of the tensor in (\ref{defphit}) and the notation $\mu_t$ for the upper index to implement the change of basis $\hat{f}^{ \mu_t}{}_{I} = \hat{f}^{ \mu}{}_{I} + ( Y{\cdot} \Im v )^{\mu} \hat{f}^{ h}{}_{I}$ as in (\ref{mutbasis}).

In the simplest case of rank two, we see that the \emph{leading-in-$t$} coefficient of the degeneration of the DHS kernels in (\ref{degfrk2}) is obtained from the degenerations of the Enriquez kernels \eqref{3.a.1} upon replacing,
\begin{align}\label{diction_gf}
     \chi_{zy}^{\vec{\mu}}(x) &\rightsquigarrow \p_x \cG_{zy}^{\vec{\mu}}(x)  \notag \\
     g^{\vec{\mu}}{}_\nu(x,y) &\rightsquigarrow f^{\vec{\mu}}{}_\nu (x,y) 
\end{align}
along with $\mu \rightsquigarrow \mu_t$ in upper indices and $h \rightsquigarrow t$ in lower indices on the left side.
We will prove in later work that the dictionary proposed above persists at all ranks, namely that the  non-separating degenerations of arbitrary DHS kernels are given~by (with $\ell \neq 1$ in the last line)
\begin{align}
\hat f^{\vec{ \mu}_t}{}_\nu (x,y) & \to \, f^{ \vmu}{}_\nu (x,y)  + \cO( t^{-1} )
\label{4.b.1aa}
\\
\hat f^{\cdots h \cdots  }{}_\nu  (x,y) & \to \,  \cO ( t^{-1}) 
\notag\\
\hat f^{\cdots h\mu_t \cdots}{}_t (x,y) & \to \,  \cO ( t^{-1})  
\notag \\
\hat f^{\vec{\mu}_t h }{}_t (x,y)
& \to \, \frac{1}{2} \, \Big( \p_x \cG^\vmu_{p_a y} (x) + \p_x \cG^\vmu_{p_b y} (x)\Big)+ \cO ( t^{-1}) 
\notag
\\
 \hat f^{\vec{ \mu}_t \vh_\ell }{}_t (x,y)&  \to \, {(2 \pi i)^{\ell-1} \over \ell  !} \,{\Ber_\ell }\,    \p_x \cG_{p_a p_b}^\vmu (x)+ \cO ( t^{-1})  
 \notag
\end{align}
The notation $\vec{\mu}_t$ refers to words in indices $\mu_{it}$ each of which is translated to a combination of $\mu_i$ and $h$ components as in the second line of (\ref{mutbasis}).

We highlight that the leading $t$ behaviour of (\ref{4.b.1aa}) precisely agrees with the non-separating degeneration \eqref{3.b.1} of Enriquez kernels, upon using the dictionary \eqref{diction_gf} between kernels and transcribing indices according to (\ref{mutbasis}) without the $Y{\cdot} \Im v$ terms. Upon inserting (\ref{4.b.1aa}) into the DHS connection \cite{DHoker:2026lgg}, its $(1,0)$ form components in $dx_i$ obey the direct analogue of the degeneration formula (\ref{5.6}) of the Enriquez connection where the $\chi^{\vI}(x_i,p_{a,b})$ in the series ${\bf X}$ translates into $\p_x \cG^{\vI}(x_i,p_{a,b})$ via (\ref{diction_gf}).\footnote{The components of $b_{iI}$, $a_i^{I}$ entering the residues $t_{ia}$, $t_{ib}$ of (\ref{5.7}) also need to be adapted to the change of basis (\ref{mutbasis}) towards term-by-term single-valuedness in $p_a$, $p_b$.}

By the modular properties of DHS kernels, their $\mA$-cycle degenerations in (\ref{4.b.1aa}) determine their behaviour under the degeneration of an arbitrary homologically non-trivial cycle. The relations between Enriquez and DHS kernels in \cite{DHoker:2025szl} can then be used to infer the more challenging $\mB$-cycle degenerations of Enriquez kernels where a Laurent polynomial
structure similar to that of $B$-elliptic multiple zeta values \cite{Enriquez:Emzv, Broedel:2018izr, Zerbini:2017usf} is expected and which will be interesting to compare with the $1/t$ corrections of (\ref{4.b.1aa}) to be studied in upcoming work.

\section{Conclusions}

In this work, we have determined the behaviour of Enriquez and DHS integration kernels
for Riemann surfaces of arbitrary genus $h$ under non-separating degenerations.
The explicit results \eqref{3.b.1} and \eqref{4.b.1aa} are entirely expressible in terms of Enriquez and DHS
kernels of genus $h{-}1$ and introduce two punctures into the degeneration of associated multi-variable connections as showcased by the intuitive form \eqref{5.6} in the Enriquez case.
The same closure property is expected for separating degenerations of the surface as exemplified
by DHS kernels at genus two \cite{DHoker:2023vax} and Enriquez kernels at arbitrary genus in the maximal separating degeneration
\cite{Ichikawa:2025kbi}.

Our results lend themselves to a plethora of applications of interest to mathematics and physics. 

In number theory and algebraic geometry, our degeneration formulas provide crucial stepping stones to investigating the behaviour of configuration-space periods on Riemann surfaces at the boundary of moduli space including modular graph tensors \cite{Kawazumi:lecture, Kawazumi, DHoker:2020uid} and higher-genus generalizations of elliptic associators \cite{EnriquezEllAss, Gonzalez:2020, DHoker:2023vax, Tani:2025} and elliptic multiple zeta values \cite{Enriquez:Emzv, Broedel:2015hia, Lochak:2020, Baune:2025sfy}. In particular, the characteristic form (\ref{5.7}) of the Lie algebra generators for the additional punctures in the degenerate connections is expected to inform the appearance of multiple zeta values in the asymptotics of non-holomorphic modular tensors as it was the case for equivariant iterated Eisenstein integrals at genus~1 \cite{Brown:2017qwo2, Dorigoni:2024oft, Dorigoni:2024iyt}.

In string amplitudes, DHS and Enriquez kernels are key ingredients of the conformal field theory correlators in their moduli-space integrand \cite{DHoker:2023khh, DHoker:2025jgb}. Our degenerations feed into bootstrap constructions of as-of-yet unknown correlators in the ambitwistor-string inspired approach of \cite{Geyer:2021oox, Geyer:2024oeu, Monteiro:2025qai} and integration techniques to determine the low-energy expansion.

For Feynman integrals in particle physics and gravity, Riemann surfaces are parts of the growing zoo of hidden geometries \cite{Weinzierl:2022eaz, Bourjaily:2022bwx, Marzucca:2023gto,Duhr:2024uid} and also carry crucial information on their higher-dimensional varieties. The degenerate connections of this work will be instrumental in successively solving the differential equations of Feynman integrals in one kinematic variable at a time.

\section*{Acknowledgments}

We are grateful to Guillaume Bossard, Benjamin Enriquez and Federico Zerbini for valuable discussions, to Martijn Hidding for participation in early stages of this project, and to Leila Schneps for crucial input concerning the appearance of Bernoulli series upon degeneration. All of us would like to thank the Erwin Schr\"odinger International Institute for Mathematics and Physics (ESI), University of Vienna (Austria), for the opportunity to participate in the Thematic Programme ``Amplitudes and Algebraic Geometry'' in 2026 where a significant part of this work has been accomplished and for the support given. MS would also like to thank Nordita (Sweden) for its kind hospitality while part of this work was being carried out. The research of ED was supported in part by NSF grant PHY-22-09700. The research of MB, AK, MS and OS is funded by the European Union under ERC Synergy Grant MaScAmp 101167287. Views and opinions expressed are however those of the author(s) only and do not necessarily reflect those of the European Union or the European Research Council. Neither the European Union nor the granting authority can be held responsible for them. 


\providecommand{\href}[2]{#2}\begingroup\raggedright\endgroup

%


\onecolumngrid 

\appendix

\begin{center}
\textbf{\large Appendix}
\end{center}

\setcounter{equation}{0} 
\renewcommand{\theequation}{A\arabic{equation}} 

In this appendix, we shall prove the general formulas (\ref{3.b.1}) for the non-separating degeneration of Enriquez kernels.
The proof of each one of the five relations in (\ref{3.b.1}), which we repeat here for convenience,
\begin{subequations}
\label{eq:degapp}
\begin{align}
\widehat g^{\vmu}{}_\nu (x,y) & \to \, g^{ \vmu}{}_\nu (x,y) 
\label{a}
\\
\widehat g^{\cdots h \cdots }{}_\nu  (x,y) & \to \, 0 
\label{b} \\
\label{c}
\widehat g^{\cdots h \mu \cdots }{}_h (x,y) & \to \, 0 
\\
\widehat g^{\vmu h }{}_h (x,y)
& \to \, \frac{1}{2}\Big(  \chi^\vmu _{p_a y} (x) + \chi^\vmu_{p_b y} (x)\Big)
\label{d} 
\\
   \widehat g^{\vmu \, \vh_\ell }{}_h (x,y)&  \to \, {(2 \pi i)^{\ell-1} \over \ell! } \, {\Ber_\ell }\,   \chi_{p_a p_b}^\vmu (x)
   \quad \ \ \ell \neq 1
\label{e}
\end{align}
\end{subequations}
proceeds from the recursive relation (\ref{2.7}) which systematically  relates every Enriquez kernel of rank $r{+}2$ to kernels of rank less than or equal to $r{+}1$ for $r\geq 1$. Therefore, the proof of the relations in (\ref{3.b.1}) will proceed by induction on the total rank $R$ of the Enriquez kernels, starting at rank $R=2$ with the results given in (\ref{3.a.1}) and at rank $R=3$ given in (\ref{3.a.5}). One straightforwardly verifies that the cases for both ranks two and three agree with the general expression given in (\ref{3.b.1}) so that the initial condition of the proof by induction is satisfied.

\subsection*{Degeneration of $\widehat g^\vmu{}_\nu(x,y)$}
We start by proving \eqref{a}. Following the recursion (\ref{2.7}), we have, 
\begin{align}\label{gamunu}
  \widehat{g}^{\alpha \vec{\mu}}{}_\nu(x,y)& =  - \oint _{\mA^\alpha} dt\,  \widehat{g}^{ \beta }{}_\nu  (x,t) \, \widehat{g}^{\vec{\mu}} {}_\beta  (t,y)- \oint _{\mA^\alpha}  dt\,\widehat{g}^{ h}{}_\nu  (x,t)  \, \widehat{g}^{\vec{\mu} } {}_{h}  (t,y)   \notag \\
  &\quad  -  \sum_{k = 1}^{r-1} (-2 \pi i)^k { \Ber_k \over k!} \, \delta ^{\mu_1 \cdots \mu_k }_\alpha \, \widehat{g}^{\alpha \mu_{k + 1} \cdots \mu_r}{}_\nu(x,y)   -    \widehat{\om}_\nu(x)   (-2 \pi i)^{r+1} { \Ber_{r+1} \over r!} \, \delta ^{\alpha \vec{ \mu}}_\nu \notag \\
  & \rightarrow - \oint _{\mA^\alpha}dt\,  g^{ \beta }{}_\nu  (x,t)  \, g^{\vec{\mu}} {}_\beta  (t,y) -  \sum_{k = 1}^{r-1} (-2 \pi i)^k { \Ber_k \over k!} \, \delta ^{\mu_1 \cdots \mu_k }_\alpha \, g^{\alpha \mu_{k + 1} \cdots \mu_r}{}_\nu(x,y) -   \om_\nu(x)   (-2 \pi i)^{r+1} { \Ber_{r+1} \over r!} \, \delta ^{\alpha \vec{ \mu}}_\nu \notag \\
  & =  g^{\alpha \vec{\mu}}{}_\nu(x,y)
\end{align}
where we have used the fact that $\widehat{g}^{ h}{}_\nu \rightarrow 0$ by the initial condition (\ref{3.a.1}), and the third line is simply the recursion (\ref{2.7}) for the genus $h{-}1$ Enriquez kernel.

\subsection*{Degeneration of $\widehat{g}^{\cdots h \cdots}{}_{\mu}(x,y)$}
Next, we prove \eqref{b}. Here we need to distinguish whether the first index is $h$ or not. In the second case, we find
\begin{align}
  \widehat{g}^{\mu \vI } {}_\nu(x,y)& =  - \oint _{\mA^{\mu}} dt\, \widehat{g}^{ \beta }{}_\nu  (x,t)  \, \widehat{g}^{\vI} {}_\beta  (t,y)- \oint _{\mA^{\mu}} dt\, \widehat{g}^{ h}{}_\nu  (x,t)  \,\widehat{g}^{\vI} {}_{h}  (t,y)   \notag \\
  &\quad  -  \sum_{k = 1}^{r-1} (-2 \pi i)^k { \Ber_k \over k!} \, \delta ^{I_1 \cdots I_k }_{\mu} \, \widehat{g}^{\mu I_{k+1} \cdots I_r}{}_\nu(x,y)   -    \widehat{\om}_\nu(x)   (-2 \pi i)^{r+1} { \Ber_{r+1} \over r!} \, \delta ^{\mu I_1 \cdots  I_r }_\nu \notag \\
  & \rightarrow  0
\end{align}
where we used the fact that $\widehat{g}^{ h}{}_\nu   \rightarrow 0$ to cancel the second term. Since at least one index among $I_1 \cdots I_r$ is $h$, the Kronecker delta in the fourth term vanishes, and so does the degeneration of $\widehat{g}^{I_1 \cdots   I_r} {}_\beta  (t,y)$ in the first term by the inductive assumption. For the third term, each summand $\delta ^{I_1 \cdots I_k }_{\mu} \, \widehat{g}^{\mu I_{k+1} \cdots I_r}{}_\nu(x,y)$ with respect to $k$ has a vanishing degeneration, either 
because of the Kronecker delta or because of the vanishing Enriquez-kernel degeneration by the inductive assumption.

Similarly, if the first index is $h$:
\begin{align}
  \widehat{g}^{h \vI} {}_\nu(x,y)& =  - \oint_{\mA^{h }} dt\,  \widehat{g}^{ \beta }{}_\nu  (x,t)  \, \widehat{g}^{\vI} {}_\beta  (t,y)- \oint_{\mA^{h}} dt \,\widehat{g}^{ h}{}_\nu  (x,t) \, \widehat{g}^{\vI} {}_{h}  (t,y)   \notag \\
  &\quad  -  \sum_{k = 1}^{r-1} (-2 \pi i)^k { \Ber_k \over k!} \, \delta ^{I_1 \cdots I_k }_{h} \, \widehat{g}^{h I_{k+1} \cdots I_r}{}_\nu(x,y)   -    \widehat{\om}_\nu(x)   (-2 \pi i)^{r+1} { \Ber_{r+1} \over r!} \, \delta ^{h I_1 \cdots  I_r }_\nu \notag \\
  & \rightarrow  \oint_{B_\epsilon(p_a)} dt\, g^{ \beta }{}_\nu  (x,t) \,{\rm deg} \big[ \widehat{g}^{\vI} {}_\beta  (t,y) \big]  = 0 
  \label{lstline}
\end{align}
where the second, third and fourth term vanish by the degenerations $\widehat{g}^{ h}{}_\nu  (x,t) \rightarrow 0$ and $\widehat{g}^{h I_{k+1} \cdots I_r}{}_\nu(x,y) \rightarrow 0$ (using the inductive assumption) and by $\delta ^{h I_1 \cdots  I_r }_\nu=0$. In the last line and throughout this appendix, ${\rm deg} \big[ \widehat{g}^{\vI} {}_\beta  (t,y) \big]$ instructs us to insert the degeneration of $\widehat{g}^{\vI} {}_\beta  (t,y)$ (which is known by the inductive assumption), and we write $B_\epsilon(p_a)$ to identify $\mA^h$ as a circle of infinitesimal radius around $p_a$ with negative orientation (which explains the minus sign relative to the first term on the right side of the first line). The integral in the last line of (\ref{lstline}) vanishes regardless of whether the indices of $\vI$ are $\vec{\mu}$ or $h$: By the inductive assumption, all components of the form $\widehat{g}^{\vI} {}_\beta  (t,y)$ give rise to non-singular integrands throughout $t\in B_\epsilon(p_a)$.

 \subsection*{Degeneration of $\widehat{g}^{\cdots h\mu\cdots}{}_{h}(x,y)$}
We now prove \eqref{c}. Proceeding similarly to the proof of \eqref{b}, we distinguish the two cases whether the first index is $h$ or not. In the second case, we have (with $I_1\cdots I_r$ containing a sequence $\cdots h\alpha \cdots$) 
\begin{align}
  \widehat{g}^{\mu \vI } {}_{h}(x,y)& =  - \oint _{\mA^{\mu}}  dt\,\widehat{g}^{ \beta }{}_{h}  (x,t)  \, \widehat{g}^{\vI} {}_\beta  (t,y)- \oint _{\mA^{\mu}} dt\,  \widehat{g}^{ h}{}_{h} (x,t) \, \widehat{g}^{\vI} {}_{h}  (t,y)   \notag \\
  &\quad  -  \sum_{k = 1}^{r-1} (-2 \pi i)^k { \Ber_k \over k!} \, \delta ^{I_1 \cdots I_k }_{\mu} \, \widehat{g}^{\mu I_{k+1} \cdots I_r}{}_{h}(x,y)  -    \widehat{\om}_{h}(x)   (-2 \pi i)^{r+1} { \Ber_{r+1} \over r!} \, \delta ^{\mu \vI }_{h} \notag \\
  & \rightarrow 0 
\end{align}
The first term vanishes because the non-separating degeneration of $\widehat{g}^{\vI} {}_\beta  (t,y) $ does (one of the $I_j$ is $h$), and the fourth term vanishes with $\delta ^{\mu \vI }_{h} = 0$. The second term vanishes by the inductive assumption since $g^{ \vI }{}_{h}$ contains a sequence $g^{ \cdots h\alpha \cdots }{}_{h}$, and the third term has a vanishing factor in each summand $\delta ^{I_1 \cdots I_k }_{\mu}  \, \widehat{g}^{\mu I_{k+1} \cdots I_r}{}_{h}(x,y)$ (by the same use of the inductive assumption if $\delta ^{I_1 \cdots I_k }_{\mu} =1$).

The second case is when the first index is $h$ (such that $I_1\cdots I_r$ must contain at least one index $\mu$):
\begin{align}
  \widehat{g}^{h \vI } {}_{h}(x,y)& =  - \oint _{\mA^{h}} dt\, \widehat{g}^{ \beta }{}_{h}  (x,t)  \, \widehat{g}^{\vI} {}_\beta  (t,y)- \oint _{\mA^{h}} dt\,  \widehat{g}^{ h}{}_{h}   (x,t)  \, \widehat{g}^{\vI } {}_{h}  (t,y)   \notag \\
  & \quad -  \sum_{k = 1}^{r-1} (-2 \pi i)^k { \Ber_k \over k!} \, \delta ^{I_1 \cdots I_k }_{h} \, \widehat{g}^{h I_{k+1} \cdots I_r}{}_{h}(x,y)  -    \widehat{\om}_\nu(x)   (-2 \pi i)^{r+1} { \Ber_{r+1} \over r!} \, \delta ^{h \vI}_{h} \notag \\
  & \rightarrow  \oint_{B_\epsilon(p_a)} dt\,   g^{ \beta }{}_{h}  (x,t) \,{\rm deg}\big[ \widehat{g}^{ \vI} {}_\beta  (t,y) \big]  + \oint _{B_\epsilon(p_a)} dt\,  g^{ h}{}_{h}(x,t)  \, {\rm deg} \big[ \widehat{g}^{ \vI } {}_{h}  (t,y)  \big] = 0
\end{align}
The Kronecker delta of the fourth term and the degeneration of each summand $\delta ^{I_1 \cdots I_k }_{h}$ $\widehat{g}^{h I_{k+1} \cdots I_r}{}_{h}(x,y) $ of the third term vanish since one of $I_j$ is $\mu$. The first integral in the third line vanishes either because the integrand ${\rm deg}\big[ \widehat{g}^{ I_1 \cdots   I_r} {}_\beta  (t,y) \big]$ is zero, or, in the case where all indices are taken to be $\vec{\mu}$, because it is regular throughout $t\in B_\epsilon(p_a)$. Similarly for the second integral in the third line, the integrand ${\rm deg} \big[ \widehat{g}^{ I_1 \cdots   I_r } {}_{h}  (t,y)  \big]$ with $r\geq 1$ and at least one index $I_j = \mu$ can never produce the singular $\chi(t,p_a)$ in its degeneration by inspecting the cases~\eqref{eq:degapp} at lower rank, so the integral over $B_\epsilon(p_a)$ vanishes (recall that $\chi^\vmu(t,p_a)$ with $\vmu \neq \emptyset $ is non-singular).

 \subsection*{Degeneration of $\widehat{g}^{\alpha \vec{\mu}h}{}_{h} (x,y)$}
Let us now prove \eqref{d}. We have (also in case of empty $\vec{\mu}$)
\begin{align}
  \widehat{g}^{\alpha \vec{\mu} h}{}_{h}(x,y)& =  - \oint _{\mA^\alpha} dt\,  \widehat{g}^{ \beta }{}_{h}  (x,t) \,   \widehat{g}^{\vec{\mu}  h} {}_\beta  (t,y)- \oint _{\mA^\alpha} dt\,  \widehat{g}^{ h}{}_{h}  (x,t) \,  \widehat{g}^{\vec{\mu}  h } {}_{h}  (t,y) \notag  \\
  &\quad  -  \sum_{k = 1}^{r} (-2 \pi i)^k { \Ber_k \over k!} \, \delta ^{\mu_1 \cdots \mu_k }_{ \alpha}\,  \widehat{g}^{\alpha \mu_{k + 1} \cdots \mu_r h}{}_{h}(x,y)  -   \widehat{\om}_{h}(x)   (-2 \pi i)^{r+1} { \Ber_{r+1} \over r!} \, \delta ^{\alpha \vec{ \mu} h}_{h}\notag \\
  & \rightarrow   - \frac{1}{4} \oint _{\mA^\alpha}dt\, \big(  \chi_{p_a t} (x) + \chi_{p_b t} (x)\big)\, \big(  \chi^\vmu _{p_a y} (t) + \chi^\vmu_{p_b y} (t)\big)  \notag \\
  &\quad - \frac{1}{2}  \sum_{k = 1}^{r} (-2 \pi i)^k { \Ber_k \over k!}  \delta ^{\mu_1 \cdots \mu_k }_{ \alpha} \, \big(  \chi^{\alpha \mu_{k + 1} \cdots \mu_r}_{p_a y} (x) + \chi^{\alpha \mu_{k + 1} \cdots \mu_r}_{p_b y} (x)\big)  \notag  \\
  &  = \frac{1}{2} \oint _{\mA^\alpha} dt\,  \chi (x,t) \, \big(  \chi^\vmu _{p_a y} (t) + \chi^\vmu_{p_b y} (t) \big)  \notag  - \frac{1}{2}  \sum_{k = 1}^{r} (-2 \pi i)^k { \Ber_k \over k!}  \delta ^{\mu_1 \cdots \mu_k }_{ \alpha}\, \big(  \chi^{\alpha \mu_{k + 1} \cdots \mu_r}_{p_a y} (x) + \chi^{\alpha \mu_{k + 1} \cdots \mu_r}_{p_b y} (x)\big)  \notag  \\
  & =\frac{1}{2} \,\big( \chi^{\alpha \vmu} _{p_a y} (x) + \chi^{\alpha\vmu}_{p_b y} (x) \big) 
  \label{gamh1h1}
\end{align}
where we have used the fact that $ \widehat{g}^{\vec{\mu}  h } {}_\beta\rightarrow  0 $ in the first line and the inductive assumption in passing to the third line. The terms of the third line involving $\chi(x,p_a)$, $\chi(x,p_b)$ vanish as a consequence of the vanishing period of $\chi^\vmu _{p_a y} (t) + \chi^\vmu_{p_b y} (t)$. The last line follows from the convolution formula \eqref{2.7a}.

\subsection*{Degeneration of $\widehat{g}^{\alpha\vec{\mu}}{}_{h}(x,y)$}
Let us now consider $\widehat{g}^{\alpha\vec{\mu}}{}_{h}(x,y)$. This is a special case of \eqref{e} when $\ell=0$ that needs to be treated separately: 
\begin{align}
  \widehat{g}^{\alpha \vec{\mu}}{}_{h}(x,y)& =  - \oint _{\mA^\alpha} dt\,   \widehat{g}^{ \beta }{}_{h}  (x,t) \,  \widehat{g}^{\vec{\mu}} {}_\beta  (t,y)- \oint _{\mA^\alpha} dt\,  \widehat{g}^{ h}{}_{h}  (x,t) \,  \widehat{g}^{\vec{\mu} } {}_{h}  (t,y) \notag \\
  &\quad  -  \sum_{k = 1}^{r-1} (-2 \pi i)^k { \Ber_k \over k!} \, \delta ^{\mu_1 \cdots \mu_k }_{ \alpha}\,  \widehat{g}^{\alpha \mu_{k + 1} \cdots \mu_r}{}_{h}(x,y)   -   \widehat{\om}_{h}(x)   (-2 \pi i)^{r+1} { \Ber_{r+1} \over r!} \, \delta ^{\alpha \vec{ \mu}}_{h}\notag \\
  & \rightarrow \frac{1}{2\pi i}
  \bigg({-} \oint _{\mA^\alpha} dt\, \chi_{p_ap_b}^{\beta }(x)  g^{\vec{\mu}} {}_\beta  (t,y)-  \frac{1}{2} \oint _{\mA^\alpha}   dt\, \big(  \chi_{p_a t}(x)+  \chi_{p_b t}(x) \big) \chi_{p_ap_b}^{\vec{\mu}}(t) \bigg) \notag \\
  &\quad -  \frac{1}{2\pi i} \sum_{k = 1}^{r-1} (-2 \pi i)^k { \Ber_k \over k!} \, \delta ^{\mu_1 \cdots \mu_k }_{\alpha}\, \chi_{p_ap_b}^{\alpha \mu_{k+1} \cdots \mu_r}(x)   \notag \\
  & =  \frac{1}{2\pi i}  \bigg({-}  \chi_{p_ap_b}^{\beta}(x)  \delta^{\alpha}_\beta \delta^{\vec{\mu}}_\beta(-2\pi i)^r \frac{\Ber_r}{r!} + \oint _{\mA^\alpha} dt\, \chi(x,t)\, \chi_{p_ap_b}^{\vec{\mu}}(t)  - \sum_{k = 1}^{r-1} (-2 \pi i)^k { \Ber_k \over k!} \, \delta ^{\mu_1 \cdots \mu_k }_{\alpha}\, \chi_{p_ap_b}^{\alpha \mu_{k+1} \cdots \mu_r}(x) \bigg) \notag \\
  & =  \frac{1}{2\pi i}  \bigg( \oint _{\mA^\alpha} dt\, \chi(x,t)\, \chi_{p_ap_b}^{\vec{\mu}}(t) - \sum_{k = 1}^{r} (-2 \pi i)^k { \Ber_k \over k!} \, \delta ^{\mu_1 \cdots \mu_k }_{\alpha}\, \chi_{p_ap_b}^{\alpha \mu_{k+1} \cdots \mu_r}(x) \Big) = \frac{1}{2\pi i} \, \chi_{p_ap_b}^{\alpha \vec{\mu}}(x) 
\end{align}
where in the third line we used the inductive assumption, then in fourth line we have used the fact that $\chi_{p_ap_b}^\vmu(t)$ has vanishing period as well as the  period formula \eqref{2.5a}, and in the last line one recognises the convolution formula \eqref{2.7a}.\footnote{Note that there is an implicit summation over $\beta$, but not over $\alpha$. In other words, $\chi_{p_a p_b}^{\beta}(x)  \delta^{\alpha}_\beta \delta^{\vec{\mu}}_\beta =\chi_{p_a p_b}^{\alpha}(x)  \delta^{\vec{\mu}}_\alpha$, where on the right hand side no summation over $\alpha$ is implied.}

 \subsection*{Degeneration of $\widehat{g}^{\vmu \vec{h}_\ell}{}_{h} (x,y)$}
Let us finally prove \eqref{e} for all values  $\ell\geq 2$, and we will initially focus on the case with no vector index in $\{1,\cdots,h{-}1 \}$, i.e.\ $\vec{\mu} = \emptyset$.

For the inductive step (in the number of upper indices equal to $h$, still keeping $\vec{\mu} = \emptyset$), we distinguish even and odd values of $\ell$:
\begin{itemize}
\item odd values of $\ell \geq 3$:
\begin{align}
  \widehat{g}^{\vec{h}_\ell}{}_{h}(x,y)& =  - \oint _{\mA^{h}} dt\,  \widehat{g}^{ \beta }{}_{h}  (x,t) \,  \widehat{g}^{\vec{h}_{\ell-1} } {}_\beta  (t,y)- \oint _{\mA^{h}} dt\,  \widehat{g}^{ h}{}_{h}  (x,t) \,   \widehat{g}^{\vec{h}_{\ell-1} } {}_{h}  (t,y)      \notag \\
  &\quad -  \sum_{k = 1}^{\ell-2} (-2 \pi i)^k { \Ber_k \over k!} \, \widehat{g}^{\vec{h}_{\ell-k}}{}_{h}(x,y)   -   \widehat{\om}_{h}(x)   (-2 \pi i)^{\ell} { \Ber_{\ell} \over (\ell{-}1)!}  \notag \\
  & \rightarrow  - \frac{1}{2} (2 \pi i)^{\ell-2} { \Ber_{\ell-1} \over (\ell{-}1)!}  \oint_{B_\epsilon(p_a)} dt\,\big(  \chi_{p_a t}(x)+  \chi_{p_b t}(x) \big) \, \chi_{p_ap_b} (t) + 2 \pi i \Ber_1  (2 \pi i)^{\ell-2} { \Ber_{\ell-1} \over (\ell{-}1)!} \chi_{p_ap_b}(x) \, \notag \\
& =  \pi i (2 \pi i)^{\ell-2} { \Ber_{\ell-1} \over (\ell{-}1)!} \chi_{p_ap_b}(x) - \pi i   (2 \pi i)^{\ell-2} { \Ber_{\ell-1} \over (\ell{-}1)!} \chi_{p_ap_b}(x) =0
\end{align} 
where in the second line we used  $\Ber_{\ell}=0$ for the odd $\ell$ under consideration and $\widehat{g}^{\vec{h}_{\ell-1} } {}_\beta  \rightarrow 0$ for arbitrary $\ell\geq 0$. The sum over $k$ in the first line has $k=1$ as its only non-zero term
(with $\Ber_1 = -\frac{1}{2}$) since $\Ber_k$ vanishes for $k\geq 3$ odd and the degeneration of $\widehat{g}^{\vec{h}_{\ell-k}}{}_{h}(x,y)$ vanishes for $k\geq 2$ even by the inductive assumption. Finally, the last line follows from:
\begin{equation}
  \frac{1}{2}  \oint_{B_\epsilon(p_a)} dt\,\big(  \chi_{p_a t}(x)+  \chi_{p_b t}(x) \big) \, \chi_{p_ap_b} (t) =  -\oint_{B_\epsilon(p_a)} dt\, \chi (x,t) \, \chi_{p_ap_b} (t)  = - \pi i \chi_{p_a p_b} (x)
\end{equation}

\item even values of $\ell \geq 4$:
\begin{align}
 \widehat{g}^{\vec{h}_{\ell}}{}_{h}(x,y)& =  - \oint _{\mA^{h}}  dt\, \widehat{g}^{ \beta }{}_{h}  (x,t) \,   \widehat{g}^{\vec{h}_{\ell-1} } {}_\beta  (t,y)- \oint _{\mA^{h}} dt\,  \widehat{g}^{ \vec{h}_{\ell-1}}{}_{h}  (x,t) \,  \widehat{g}^{\vec{h}_{\ell-1}} {}_{h}  (t,y)  \notag \\
 &\quad -  \sum_{k = 1}^{\ell-2} (-2 \pi i)^k { \Ber_k \over k!} \, \widehat{g}^{\vec{h}_{\ell-k}}{}_{h}(x,y)   -   \widehat{\om}_{h}(x)   (-2 \pi i)^{\ell} { \Ber_{\ell} \over (\ell{-}1)!}  \notag \\
  & \rightarrow   \bigg( {-}  \sum_{k = 1}^{\ell-2} (-2 \pi i)^k { \Ber_k \over k!}(2 \pi i)^{\ell-k-1} { \Ber_{\ell-k} \over (\ell{-}k)!}   -     (2 \pi i)^{\ell-1} { \Ber_{\ell} \over (\ell{-}1)!}  \bigg)\chi_{p_ap_b}(x) \notag \\
  &=(2 \pi i)^{\ell-1} \, { \Ber_{\ell} \over \ell!}\,  \chi_{p_ap_b}(x) 
\end{align}
where the first two terms in the first line have vanishing degenerations (both $ \widehat{g}^{\vec{h}_{\ell} } {}_\beta  (t,y)$ and, by the inductive assumption, $\widehat{g}^{\vec{h}_{\ell-1} } {}_{h}  (t,y)$), while the last two terms are both proportional to $\chi_{p_ap_b}(x) $ by the inductive assumption.
Finally,  the sum over bilinears in Bernoulli numbers is evaluated according to the identity
\beq
\sum_{k=1}^N (-1)^k \frac{\Ber_k}{k!} \, \frac{\Ber_{N-k}}{(N{-}k)!} = - \frac{\Ber_N}{(N{-}1)!}
\eeq
among Bernoulli numbers, which can be easily proven through the rewriting
\beq
\sum_{k=0}^N (-1)^k\, \frac{\Ber_k}{k!} \, \frac{\Ber_{N-k}}{(N{-}k)!} = (1{-}N) \,\frac{\Ber_N}{N!}
\eeq
and the differential equation
\beq
\frac{-x^2}{(e^x-1)\, (e^{-x}-1)} = (1-x\partial_x)\, \frac{x}{e^x-1}
\eeq
among the associated generating functions.

This concludes our inductive  proof of \eqref{e} at empty $\vec{\mu}$. The case for arbitrary non-empty $\vec{\mu}$ can be easily proved by induction in the length $r$ of $\vec{\mu}$, following similar steps as in \eqref{gamh1h1}. For example, for even $\ell$, we have (the odd $\ell$ case is very similar):
\begin{align}
  \widehat{g}^{\alpha \vec{\mu} \vec{h}_{\ell}}{}_{h}(x,y)& =  - \oint _{\mA^\alpha} dt\,   \widehat{g}^{ \beta }{}_{h}  (x,t) \,  \widehat{g}^{\vec{\mu}  \vec{h}_{\ell}} {}_\beta  (t,y)- \oint _{\mA^\alpha}  dt\, \widehat{g}^{ h}{}_{h}  (x,t) \,  \widehat{g}^{\vec{\mu}  \vec{h}_{\ell}} {}_{h}  (t,y) \notag  \\
  &\quad  -  \sum_{k = 1}^{r} (-2 \pi i)^k { \Ber_k \over k!} \, \delta ^{\mu_1 \cdots \mu_k }_{ \alpha}\,  \widehat{g}^{\alpha \mu_{k + 1} \cdots \mu_r \vec{h}_{\ell}}{}_{h}(x,y)   -  \sum_{k = r+1}^{\ell+r-1} (-2 \pi i)^k { \Ber_k \over k!} \, \delta ^{\vec{\mu} \vec{h}_{k-r} }_{ \alpha}\,  \widehat{g}^{\alpha \vec{h}_{\ell+r-k}}{}_{h}(x,y)   \notag \\
  &\quad -   \widehat{\om}_{h}(x)   (-2 \pi i)^{r+1} { \Ber_{r+1} \over r!} \, \delta ^{\alpha \vec{ \mu} \vec{h}_{\ell}}_{h}\notag \\
  & \rightarrow   - \frac{1}{2} \oint _{\mA^\alpha}dt\, \big(\chi_{p_a t}(x)+\chi_{p_b t}(x) \big) \, \bigg((2 \pi i)^{\ell-1} { \Ber_{\ell} \over \ell!}\chi_{p_ap_b}^{\vec{\mu}}(t)  \bigg)  \notag \\
  &\quad -  \sum_{k = 1}^{r} (-2 \pi i)^k { \Ber_k \over k!}  \delta ^{\mu_1 \cdots \mu_k }_{ \alpha} \bigg( (2 \pi i)^{\ell-1} { \Ber_{\ell} \over \ell!}\chi_{p_ap_b}^{\alpha \mu_{k+1} \cdots \mu_r}(x) \bigg) \notag  \\
  & =   (2 \pi i)^{\ell-1} { \Ber_{\ell} \over \ell!} \oint _{\mA^\alpha}dt\, \chi(x,t) \,\chi_{p_ap_b}^{\vec{\mu}}(t)     -  (2 \pi i)^{\ell-1} { \Ber_{\ell} \over \ell!} \sum_{k = 1}^{r} (-2 \pi i)^k { \Ber_k \over k!}  \delta ^{\mu_1 \cdots \mu_k }_{ \alpha} \chi_{p_ap_b}^{\alpha \mu_{k+1} \cdots \mu_r}(x)  \notag \\
  & = (2 \pi i)^{\ell-1} { \Ber_{\ell} \over \ell!}\chi_{p_ap_b}^{\alpha \vec{\mu}}  (x)
\end{align}
where we have used the fact that $ \widehat{g}^{\vec{\mu}  \vec{h}_{\ell}} {}_\beta  \rightarrow  0 $ in the first line and the inductive assumption on the second and the third term in passing to the fourth line (the fourth and fifth term vanish through the Kronecker deltas). Note that we have split the sum over $k$ into the ranges $k\le r$ and $k\ge r{+}1$ in order to emphasize that all terms in the second sum vanish identically due to the Kronecker delta.
Finally, the second to last equality follows from the convolution \eqref{2.7a} and the fact that $\chi^\vmu_{p_ap_b}(t)$ has vanishing $\mA$-periods. 
\end{itemize}

\end{document}